\PassOptionsToPackage{numbers,sort&compress}{natbib}  
\documentclass[preprint,12pt]{elsarticle}
\usepackage[margin=2.5cm]{geometry}
\usepackage{enumitem}
\newlist{inlinelist}{enumerate*}{1}
\setlist*[inlinelist,1]{%
  label=(\roman*),
}
\usepackage{soul}
\usepackage{hhline}
\usepackage{subfig}
\usepackage{multirow}
\usepackage{xcolor}
\usepackage{graphicx}
\usepackage{upgreek}
\usepackage{amssymb}
\usepackage{amsfonts,amsthm,bm,amsmath} 
\usepackage{appendix}
\usepackage{times}
\usepackage{caption}
\usepackage{subfig}
\newcommand{\psubref}[1]{\protect\subref{#1}}
\usepackage{multirow}
\usepackage{xcolor}
\usepackage[linesnumbered,ruled,vlined]{algorithm2e}
\usepackage{tablefootnote}

\SetCommentSty{mycommfont}

\SetKwInput{KwInput}{Input}                
\SetKwInput{KwOutput}{Output}              

\usepackage[colorlinks]{hyperref} 
\hypersetup{ 
    colorlinks=true,       
    linkcolor=red,          
    citecolor=blue,        
    filecolor=magenta,      
    urlcolor=cyan           
}

\newcommand{\fref}[1]{Fig.~\ref{#1}}

\newcommand{\eref}[1]{Eq.~(\ref{#1})}

\newcommand{\sref}[1]{Section~\ref{#1}}
\newcommand{\tref}[1]{Table~\ref{#1}}
\setcitestyle{square}

\journal{Engineering Applications of Artificial Intelligence}
\begin{document}

\begin{frontmatter}

\title{Sequential Deep Operator Networks (S-DeepONet) for Predicting Full-field Solutions Under Time-dependent Loads}
\author[]{Junyan He$^{1}$}
\author[]{Shashank Kushwaha$^{1}$}
\author[]{Jaewan Park$^{1}$}
\author[]{Seid Koric$^{1,2}$\corref{mycorrespondingauthor}}
\cortext[mycorrespondingauthor]{Corresponding author}
\ead{koric@illinois.edu}
\author[]{Diab Abueidda$^{2,3}$}
\author[]{Iwona Jasiuk$^1$}

\address{$^1$ Department of Mechanical Science and Engineering, University of Illinois at Urbana-Champaign, Urbana, IL, USA \\
$^2$ National Center for Supercomputing Applications, University of Illinois at Urbana-Champaign, Urbana, IL, USA \\
$^3$ Civil and Urban Engineering Department, New York University Abu Dhabi, Abu Dhabi, UAE\\
}

\begin{abstract}
Deep Operator Network (DeepONet), a recently introduced deep learning operator network, approximates linear and nonlinear solution operators by taking parametric functions (infinite-dimensional objects) as inputs and mapping them to solution functions in contrast to classical neural networks that need re-training for every new set of parametric inputs. In this work, we have extended the classical formulation of DeepONets by introducing sequential learning models like the gated recurrent unit (GRU) and long short-term memory (LSTM) in the branch network to allow for accurate predictions of the solution contour plots under parametric and time-dependent loading histories. Two example problems, one on transient heat transfer and the other on path-dependent plastic loading, were shown to demonstrate the capabilities of the new architectures compared to the benchmark DeepONet model with a feed-forward neural network (FNN) in the branch. Despite being more computationally expensive, the GRU- and LSTM-DeepONets lowered the prediction error by half (0.06\% vs. 0.12\%) compared to FNN-DeepONet in the heat transfer problem, and by 2.5 times (0.85\% vs. 3\%) in the plasticity problem. In all cases, the proposed DeepONets achieved a prediction $R^2$ value of above 0.995, indicating superior accuracy. Results show that once trained, the proposed DeepONets can accurately predict the final full-field solution over the entire domain and are at least two orders of magnitude faster than direct finite element simulations, rendering it an accurate and robust surrogate model for rapid preliminary evaluations. 
\end{abstract}


\end{frontmatter}

\section{Introduction}
\label{sec:intro}
Recent technological advances in high-performance computing hardware and machine learning (ML) methods have given rise to a wide range of applications in fields like autonomous driving, image and speech recognition, bioinformatics, medical diagnosis, document categorization, and others. Physics-based modeling has shown much interest in applications of Deep Learning with Artificial Neural Networks (ANN), a branch of machine learning motivated by the brain's biological structure and operation. Without the need for expensive computing power or modeling software, a well-trained surrogate deep learning model can almost immediately produce (infer) outcomes that are comparable with traditional modeling techniques. Many data-driven surrogate deep learning models have been devised and trained to quickly solve problems in additive manufacturing \cite{wang2020machine}, topologically optimized materials and structures \cite{kollmann2020deep,he2022deep}, automatic damage detection in civil structures \citep{cha2017deep, cha2018autonomous}, bio-inspired structures \cite{kushwaha2023effect,zhang2022dynamic}, composite materials \cite{chen2019machine}, nonlinear material responses \cite{mozaffar2019deep,egli2021surrogate,abueidda2021deep}, as well as a variety of other applications. Besides data-driven models, collocation point-based physics-informed neural networks (PINN) was created by Raissi et al. \cite{raissi2019physics} and Abueidda et al. \cite{abueidda2021meshless} capable of solving partial differential equations governing deformation and stress generation in solids or other physics without the aid of finite elements or other conventional numerical techniques, other than for validation. Similar to this, Nguyen-Thanh et al. \cite{nguyen2020deep}, Samaniego et al. \cite{samaniego2020energy}, Abueidda et al. \cite{abueidda2022deep,abueidda2023enhanced}, Fuhg et al. \cite{fuhg2022mixed} and He et al. \cite{he2023deep,he2023use} devised a deep energy method (DEM), which makes use of potential energy to resolve nonlinear material responses. Besides forward problems, Haghighat et al. \cite{haghighat2021physics} used PINNs for inverse problems in solid mechanics.  Cai et al. \cite{cai2021flow} even used a combination of measured data and a physics-informed deep-learning method to obtain a solution for an ill-posed thermal fluid flow that was previously thought to be unsolvable.

Nevertheless, most of these methods require retraining or transfer learning if input parameters like loads, boundary conditions, and material properties, or geometry change. The same is true of traditional numerical methods such as finite elements (FE) in that each new input parameter value calls for a new independent simulation. To address this problem, the universal approximation theorem for operators \cite{chen1995universal} inspired Lu et al.'s Deep Operator Network \cite{lu2021learning}, often known as DeepONet, an innovative operator learning architecture. It contains two sub-networks, a branch network to encode the input functions and a trunk network to encode the input domain geometry. In its original form, both networks are feed-forward neural networks (FNNs). For a few linear and nonlinear PDEs in that landmark work, DeepONet successfully mapped between unknown parametric functions and solution spaces in addition to learning explicit operators like integrals. This gave rise to a powerful new method for solving stochastic and parametric PDEs. In the so-called physics-informed DeepONet, Wang et al. \cite{wang2021learning} improved the DeepONet formulation by including information from the governing PDE. They found improved prediction accuracy and data handling efficiency but at the expense of a higher computing cost for training. Recently, DeepONet has been used in heat conduction with a spatially variable heat source by Koric and Abueidda \cite{koric2023data}, fracture mechanics by Goswami et al. \cite{goswami2022physics}, and multiscale modeling using elastic and hyperelastic materials by Yin et al. \cite{yin2022interfacing}, and elastic-plastic stress field prediction on topologically optimized geometries \cite{he2023novel}. Although the DeepONet is capable of predicting the full-field solution, the above-mentioned works do not cover time-dependent loads. In path-dependent phenomena such as plasticity and material damage, causal relationships in time-dependent input signals are pivotal, However, a FNN in the branch network does not retain the causality of input data.

Applied loads encountered in the real world are hardly static and stationary. Often, they are time-dependent, such as wind loads, vibrations, and impact. When dealing with time-dependent signals, recurrent neural networks such as the gated recurrent unit (GRU) \cite{cho2014learning} and long short-term memory (LSTM) \cite{schmidhuber1997long} are commonly employed. They are two solutions to the vanishing gradient issues \cite{hochreiter1998vanishing} of a simple recurrent neural network. In LSTM and GRU, hidden state (memory) cells are intended to dynamically "forget" some outdated and pointless information via particular gated units that regulate the information flow inside a memory cell, preventing the multiplication of a lengthy series of numbers during temporal backpropagation. Many previous studies employed these architectures to accurately predict time-dependent phenomena such as adsorption \cite{skrobek2022implementation,skrobek2020prediction}, landslide \cite{zhang2021novel}, plastic deformation \cite{abueidda2021deep}, damage \cite{choe2021sequence}, and active noise cancellation \citep{cha2023dnoisenet,mostafavi2023deep}. Therefore, it is reasonable to consider the GRU and LSTM networks as capable candidates for encoding time-dependent input load histories. However, these recurrent neural network architectures, in their original form, are meant for predicting sequences. That is, they learn from a series of input signals and predict the corresponding output signals. Those output signals are typically 1D and only contain temporal information.  While these recurrent neural networks have been extensively used in sequence-to-sequence "translation", the application of these architectures to predict full-field, spatial contour distribution is relatively under-explored. Shi et al. \cite{shi2015convolutional} proposed a variant of LSTM known as ConvLSTM, which combines the temporal encoding capability of LSTM with the spatial encoding of convolutional neural networks. This architecture is subsequently used by Frankel et al. \cite{frankel2020prediction} to predict the stress field evolution in polycrystals. However, the ConvLSTM is limited to a structured grid due to its convolutional nature, which is not a limitation of the DeepONet model. There are also no previous studies investigating the effects of combining a recurrent neural network in the DeepONet structure.

Many nonlinear thermal, mechanical, and multiphysics analyses have time-dependent loads often coupled with highly nonlinear thermo-mechanical properties, including phase transformation and/or materially nonlinear path-dependent constitutive models, such as in plastic deformation. Therefore, capturing the time-dependent loading history is vital to accurately solving these kinds of problems. Frequently, only the final stress or temperature field solution, obtained through incremental nonlinear finite element analysis, is of interest for analysis, designs, and optimizations. Such problems can be particularly computationally challenging and time-consuming for sensitivity analysis, uncertainty quantification, topology optimization, and similar iterative procedures where thousands, or even millions of forward evaluations, are needed to be done by classical nonlinear analysis to achieve statistical convergence. Therefore, there is a scientific need for developing accurate data-driven surrogate models for time-dependent nonlinear problems to reduce the number of FE simulations and to generate full-field contours for quantities of interest. However, as discussed above, the classical DeepONets, although capable of predicting a full-field contour, lack the capability of causal relationships in input data with its FNN branch network. On the other hand, classical recurrent neural networks like GRUs and LSTMs can accurately capture time-dependent signals but are not designed to predict a full-field output with spatial information. Therefore, the objective of this work is to combine sequence learning models such as the gated recurrent unit (GRU) and long short-term memory (LSTM) in the branch network of the DeepONet. Since temporal loading histories are essentially long time series, it is reasonable to apply recurrent neural network architectures to capture the temporal information embedded in the input history. To the best of the authors' knowledge, such a combination of sequence learning models and the DeepONet branch-trunk architecture is a first in the literature and is the most significant and novel scientific contribution of this work. We have tested our approach with two example problems: (1) a transient heat conduction problem with phase transformation (solidification) and (2) a path-dependent plasticity problem, both involving significant nonlinearity and representing real-world engineering use cases. We used the proposed DeepONets to predict full-field solutions and compared performance with the classical DeepONet formulation.

This paper is organized as follows: \sref{sec:methods} introduces three neural networks architectures and provides detail on the data generation method. \sref{sec:results} presents and discusses the performance of the three models. \sref{sec:conc} summarizes the outcomes, limitations, and highlights future works.

\section{Methods}
\label{sec:methods}
\subsection{Neural network models}
\label{sec:NNs}
This work explored three different NN architectures to predict the final temperature or von Mises stress distribution. All three NNs were implemented in the DeepXDE framework \cite{lu2021deepxde} with a TensorFlow backend \cite{tensorflow2015-whitepaper}. 

\subsubsection{FNN-DeepONet}
\label{sec:don}
\begin{figure}[b!] 
    \centering
         \includegraphics[width=0.8\textwidth]{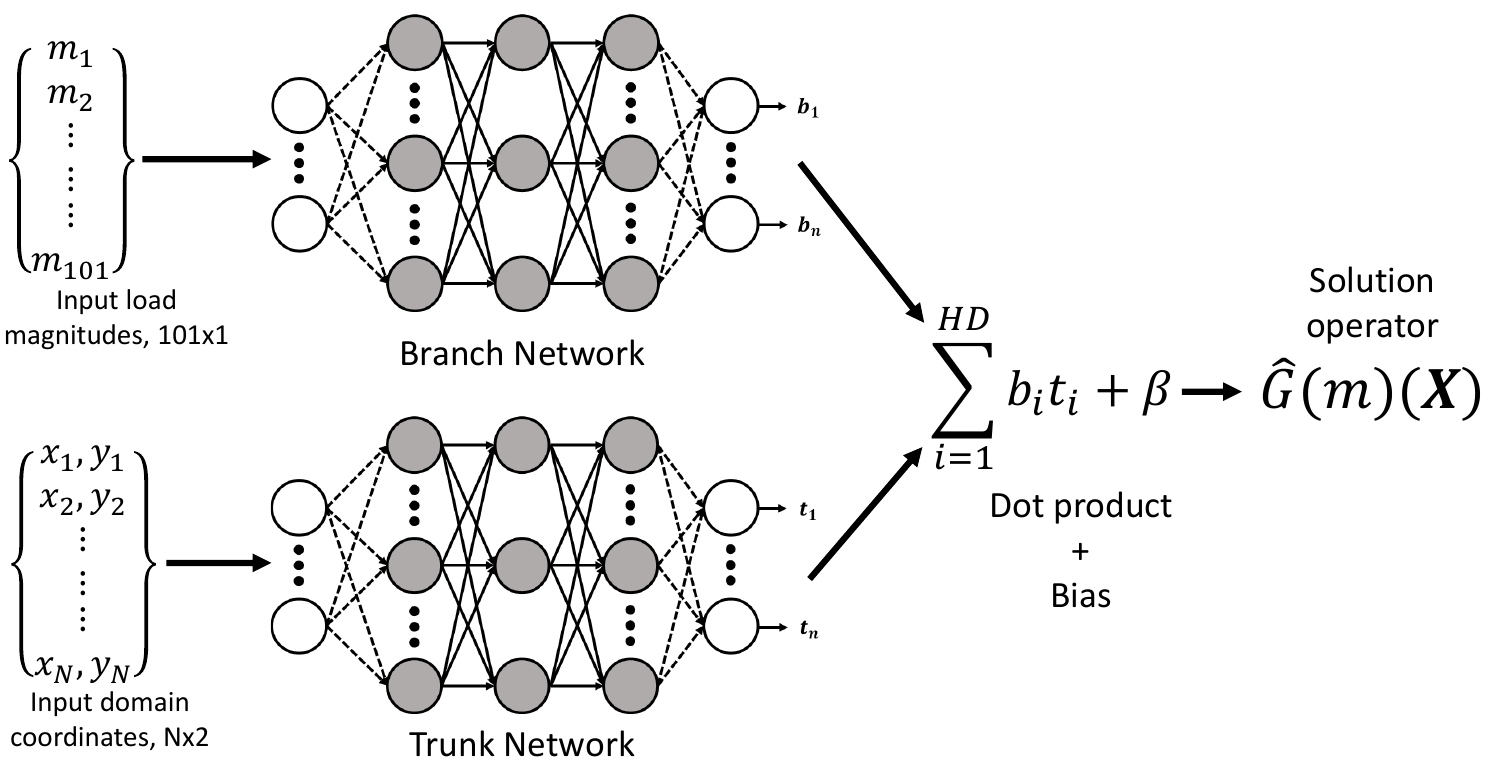}
    \caption{Schematic of the FNN-based DeepONet used in this work. $m$, $x$, $y$, $b_i$, $t_i$, $HD$, $\beta$ and $\Hat{G}$ denote the load magnitude, X coordinate, Y coordinate, branch output, trunk output, hidden dimensions, the bias vector and the approximated solution operator.}
    \label{fnn_don}
\end{figure}

A DeepONet model with FNN in both the branch and trunk networks is used as the performance baseline. In an infinite-dimension functional input space $M$, $m \in M$ represents an input function with history-dependent load magnitudes defined on $n$ control points (or input sensors) and  refers to an unknown temperature or stress field solution in the functional solution space $S$. We consider that for every input $m \in M$ there is a solution $s = s(m) \in S$ for the temperature or stress field distribution from the \sref{sec:HT} and \sref{sec:plastic_def}, which are also subject to respective boundary conditions (BCs). Consequently, the mapping solution operator of a DeepONet $G: M \to S$ can be defined as:
\begin{equation}
    G(m) = s(m).
\end{equation}
For a collection of $N$ points $\bm{X}$ on a domain, each denoted by its coordinates $(x_i,y_i)$, the DeepONet considers both the load magnitudes $m$ in its branch and positions $\bm{X}$ in its trunk and predicts the solution operator $\Hat{G}(m)(\bm{X})$ by combining intermediate encoded outputs $b_i$ (from branch) and $t_i$ (from trunk) in a dot product enhanced by bias $\beta$, as shown in a schematic of the FNN-based DeepONet model in \fref{fnn_don}. 
In a larger sense, it is possible to think of $\Hat{G}(m)(\bm{X})$ as a function of $\bm{X}$ conditioning on input $m$, and DeepONet is more general and capable than other neural networks. 

Specifically, in this work, we seek to use an FNN-DeepONet to predict the final temperature (in \sref{sec:ht_res}) and von Mises stress (in \sref{sec:plastic_res}) contours given a time-dependent input load. In both cases, the time-dependent input load function is evaluated at $n=101$ time steps to form a $101 \times 1$ input load vector $\bm{m}$, which is fed to the branch network of the FNN-DeepONet. The 2D problem geometry is described by $N$ nodes within the domain, assembled into a $N \times 2$ matrix, and fed to the trunk network of the DeepONet. The DeepONet is used as a regressor that predicts an output field of shape $N \times 1$ with the field value (e.g., temperature or Mises stress) at the end of the load step defined at each node. The outputs can be obtained via a simple forward evaluation of the DeepONet model given the above inputs. The FNN-DeepONet used in this work has seven layers in its branch and trunk networks. The numbers of neurons in the branch and trunk networks are $[101, 100, 100, 100, 100, 100, HD]$ and $[2, 100, 100, 100, 100, 100, HD]$, respectively. Here, $HD$ denotes the hidden dimension of the branch and trunk networks and was set to 100 in this work. The ReLU activation function was applied to the outputs of the branch and trunk networks of the DeepONet. This network contains 111500 trainable parameters and was used as the benchmark model to solve both problems introduced in \sref{sec:data_gen}. A larger network with a similar number of parameters as in \sref{sec:sdon} was tested but showed similar performance as this smaller network, so the smaller network was used in the result sections. The model was trained for 350000 epochs with a batch size of 64. The Adam optimizer \cite{kingma2014adam} was used and the scaled mean squared error (MSE) was used as the loss function, which is defined as:
\begin{equation}
    {\rm{MSE}} = \frac{ 1 }{ N } \sum^N_{i=1} (f_{FE} - f_{Pred})^2,
\end{equation}
where $N$, $f_{FE}$, and $f_{Pred}$ denote the number of data points, the FE-simulated field value, and the NN-predicted field value, respectively.

\subsubsection{Sequential DeepONets}
\label{sec:sdon}
The FNN-based DeepONet introduced in \sref{sec:don} uses an FNN to encode the time series signal of the input magnitude. In this work, two different recurrent neural network architectures, the GRU and LSTM, are considered in the branch network of the DeepONet, the resulting sequential DeepONets are called GRU- and LSTM-DeepONets in subsequent discussions. The new branch networks are to be designed with identical input and output signatures as the FNN branch network described in \sref{sec:don} so they can be dropped directly into the DeepONet architecture. The two sequential branch networks are described in detail below.

\begin{figure}[h!] 
    \centering
         \includegraphics[width=.85\textwidth]{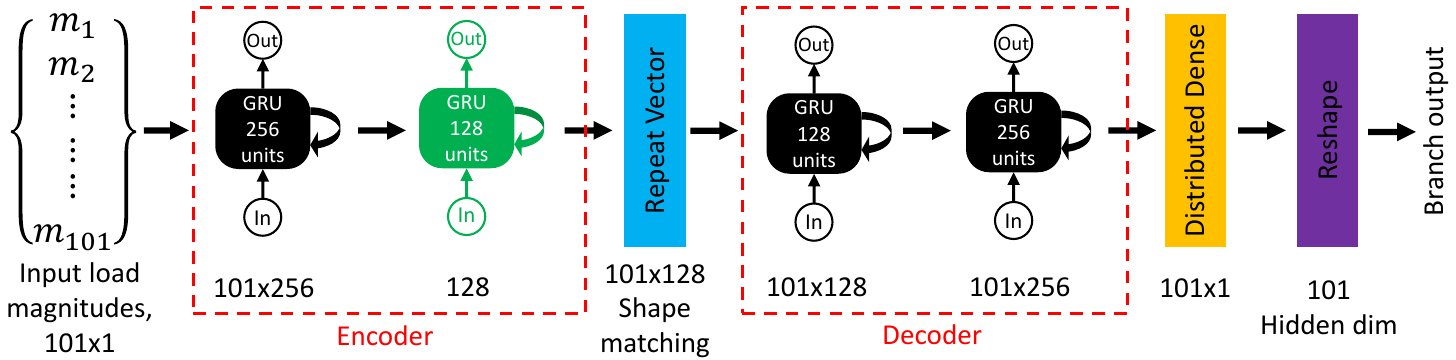}
    \caption{Schematic of the GRU branch network. The black GRU blocks return a sequence (2D outputs), while the green GRU block compresses the output into 1D. The hidden dimension for this branch network is 101, identical to the number of time steps in the input load vector.}
    \label{gru_branch}
\end{figure}

The first sequential branch network architecture considered in this work is the GRU architecture. We utilized an encoder-decoder structure inside the branch work, which was shown in many previous sequence-to-sequence prediction tasks to significantly improve the prediction accuracy \cite{malhotra2016lstm,sutskever2014sequence}. The schematic of the GRU-based branch network is shown in \fref{gru_branch}. 
The developed network consists of four GRU layers, the first two being the encoder and the last two being the decoder. The first layer of the encoder encodes the information into 256 latent features, which are compressed to a 128 vector by the second GRU layer of the encoder (green block in \fref{gru_branch}). A repeat vector layer is added to match the shape of the decoder portion of the network. The decoder portion consists of two GRU layers with 128 and 256 units, respectively, to decode the encoded upstream information. All GRU layers use a tanh activation function. Finally, a time-distributed dense layer with linear activation is used to output the results to the larger DeepONet architecture with a hidden dimension of 101. To accommodate the hidden dimension of the branch network, the trunk network of the GRU-DeepONet is an FNN with the following seven layers of neurons: $[2, 101, 101, 101, 101, 101, 101]$. The resulting GRU-DeepONet has 792422 trainable parameters.

\begin{figure}[h!] 
    \centering
         \includegraphics[width=.85\textwidth]{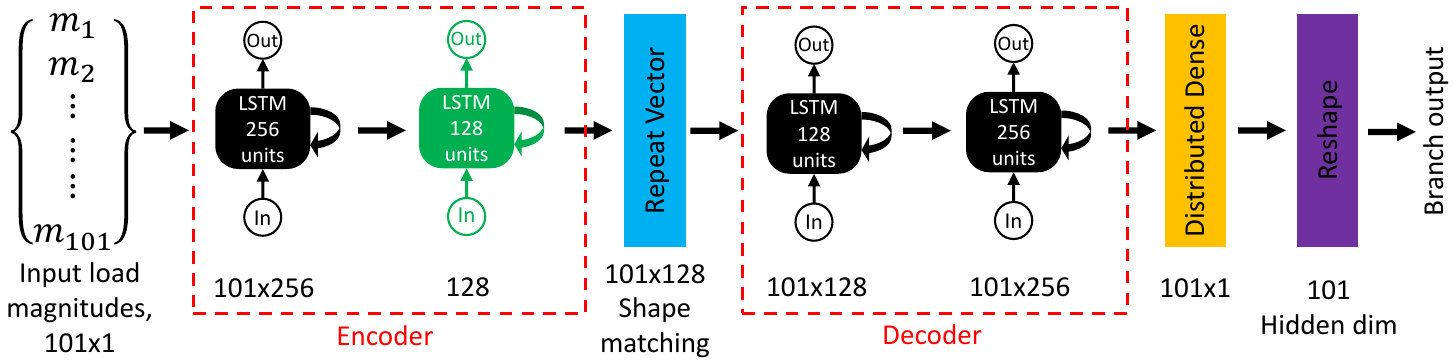}
    \caption{Schematic of the LSTM branch network. The black LSTM blocks return a sequence (2D outputs), while the green LSTM block compresses the output into 1D. The hidden dimension for this branch network is 101, identical to the number of time steps in the input load vector.}
    \label{lstm_branch}
\end{figure}

The LSTM network is also considered in the branch network of a DeepONet structure, and its schematic is shown in \fref{lstm_branch}. 
Similar to the GRU branch network, the LSTM branch network contains 4 LSTM layers, the first two form an encoder with 256 and 128 units, and the last two form a decoder with 128 and 256 units, respectively. The second LSTM block in the encoder (green block in \fref{lstm_branch}) is intended to compress the information into 1D as a 128 vector. The tanh activation function was used in all LSTM layers, and the trunk network of the LSTM-DeepONet is identical to that of the GRU-DeepONet. The proposed network has 1039206 trainable parameters. 

Identical to the FNN-DeepONet introduced \sref{sec:don}, the GRU- and LSTM-DeepONets are intended to be used as a regression model that predicts the temperature or Mises stress profiles given a time-dependent input load and a set of nodal coordinates defining the geometry. The inputs, outputs, optimizer, loss function, and training epochs are the same as those of the FNN-DeepONet.

\subsection{Data generation}
\label{sec:data_gen}
In this work, we compare the performance of the two proposed sequential DeepONets with the classical FNN-DeepONet in transient heat transfer and structural deformation problems.
\subsubsection{Transient heat transfer}
\label{sec:HT}
\begin{figure}[h!] 
    \centering
         \includegraphics[width=0.45\textwidth]{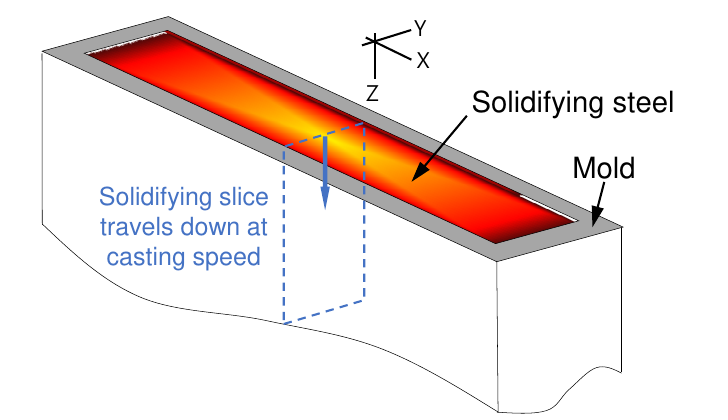}
    \caption{3D schematic of molten steel solidifying in a caster mold.}
    \label{caster}
\end{figure}

In the first example, a nonlinear, transient heat transfer problem is studied, which is representative of a solidifying shell of low-carbon steel in a continuous caster \cite{abueidda2021deep}. \fref{caster} provides a schematic to illustrate the process. 

\begin{figure}[h!] 
    \centering
         \includegraphics[width=0.9\textwidth]{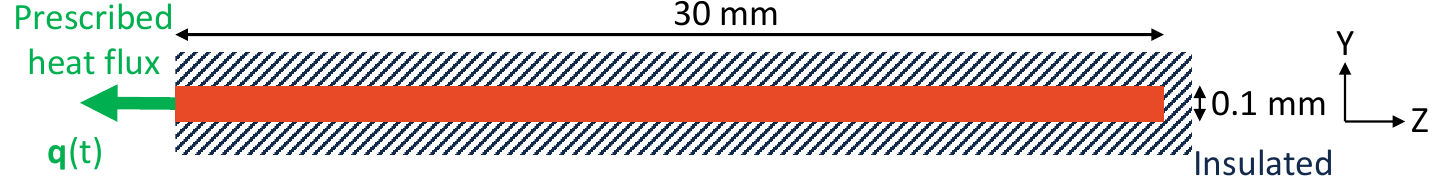}
    \caption{Schematic of the 2D domain used in the heat transfer problem.}
    \label{heat_transfer_domain}
\end{figure}

To simplify the simulation, a 2D slice cross-section of the caster is modeled, which has a length of 30 mm and a width of 0.1 mm. As the domain moves down the mold in a Lagrangian frame of reference, it is subject to a prescribed time-dependent heat flux extracting heat on one end, and all other surfaces are insulated. A schematic of the 2D problem domain is shown in \fref{heat_transfer_domain}. 
The governing equation and initial and boundary conditions of the transient heat transfer are given by:
\begin{equation}
\begin{aligned}
    \rho H(T) \frac{\partial T }{ \partial t } = \nabla \cdot \left[ k(T) \nabla T \right],\\
    -k(T) \nabla T = \bm{q}(t) , \;\; \forall \bm{x} \in \partial \Omega_q,\\
    T(x,0) = T_0,\\
\end{aligned}
\label{heat_eq}
\end{equation}
where $t$ is time, $T$ is temperature, $\rho$ is mass density, and $\bm{q}(t)$ is the time-dependent heat flux. $T_0$ = 1540 $^o$C is the uniform initial temperature. $H(T)$ and $k(T)$ denote the temperature-dependent specific enthalpy and isotropic thermal conductivity, respectively. It is highlighted that the specific enthalpy used in this work includes the latent heat effect during phase transformations, such as in solidification and transition from $\delta$-ferrite to austenite, and brings significant nonlinearity to the system \cite{Abaqus2021}. The material is a low carbon steel with a density of 7400 kg/m$^3$; other relevant material properties are shown in \tref{mat1} and \tref{mat2}.
\begin{table}[h!]
    \caption{Temperature-dependent material properties}
    \small
    \centering
    \begin{tabular}{cccc}
     Temperature [$^o$C] & \vline & Conductivity [W/(m$\cdot$K)] & Specific heat [J/(kg$\cdot$K)] \\
    \hline
    800 & \vline  &  28.934 & 695.44 \\
    1480.04 & \vline  &  34.188 & 695.44 \\
    1519.73 & \vline  &  39.000 & 824.61 \\
    \end{tabular}
    \label{mat1}
\end{table}
\begin{table}[h!]
    \caption{Latent heat properties}
    \small
    \centering
    \begin{tabular}{ccccc}
      & \vline & Latent heat [J] & Liquidus temperature [$^o$C] & Solidus temperature [$^o$C] \\
    \hline
    Value & \vline  &  245100.0 & 1524.59 & 1475.43 \\
    \end{tabular}
    \label{mat2}
\end{table}

\begin{figure}[h!] 
    \centering
     \subfloat[]{
         \includegraphics[trim={0cm 0cm 0cm .2cm},clip,width=0.4\textwidth]{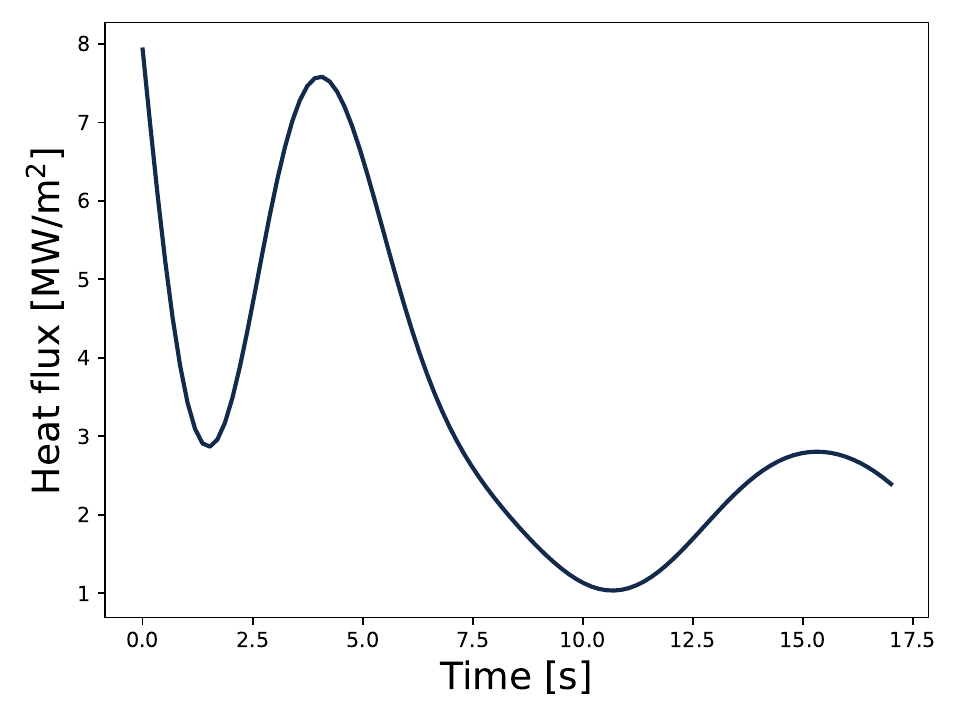}
         \label{heat_flux}
     }
     \subfloat[]{
         \includegraphics[trim={0cm 0cm 0cm .2cm},clip,width=0.4\textwidth]{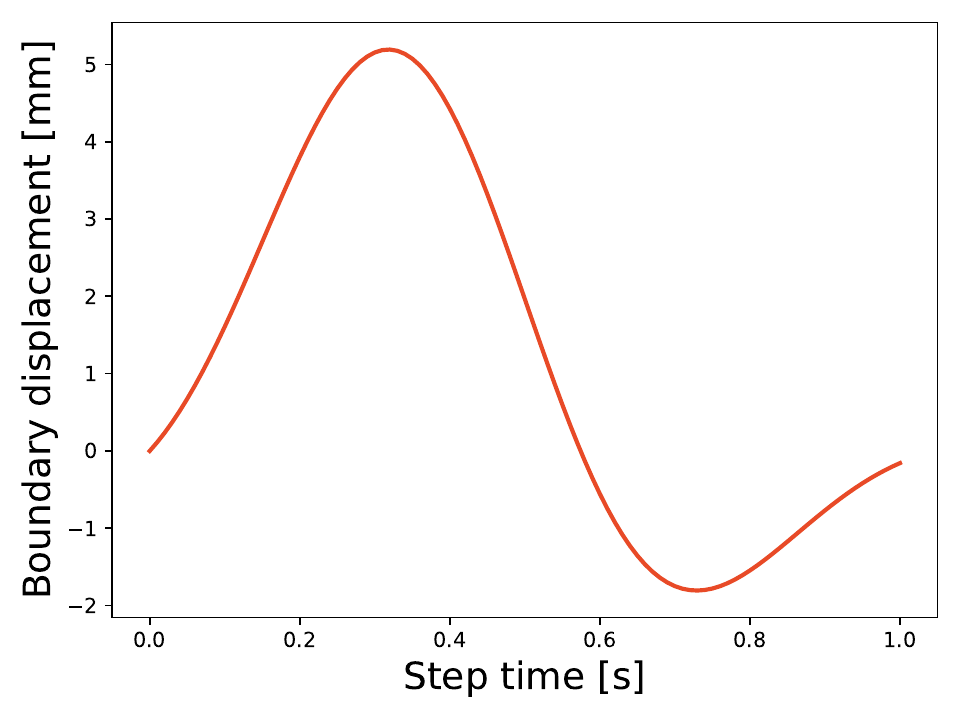}
         \label{disp}
     }
    \caption{Typical time dependent load magnitudes: \psubref{heat_flux} Boundary heat flux in the transient heat transfer problem. \psubref{disp} Applied displacement in the plastic deformation problem.}
    \label{magnitudes}
\end{figure}

The problem domain was discretized into 300 four-node bi-linear heat transfer elements (DC2D4) with an element size of 0.1 mm, and the thermal equation \eref{heat_eq} is solved for a total of 17s that the slice spends in the caster traveling down the mold by using implicit time integration in Abaqus/Standard \citep{Abaqus2021}. The temperature distribution at the end of the load step was extracted as the ground truth for NN training. To define a complex, time-dependent boundary heat flux, the sampling approach by Abueidda et al. \cite{abueidda2021deep} was used, where the time-dependent boundary heat flux is defined by six control points. The first and last control points ($_{cp}$) correspond to $t=0$ and $t=17s$, respectively. The time values ($t_{cp}$) for the four remaining control points were randomly sampled from a uniform distribution in the range $(0,17)s$. Based on experimental measurements, the flux value $q_{cp}$ generally has a decaying profile and can be approximated as:
\begin{equation}
    q_{cp} = A( t_{cp} + 1 )^{-B} + C,
\end{equation}
where $A \in [3,8]$, $B \in [0.3,0.7]$, and $C \in [-0.5, 0.5]$ were randomly chosen variables from their respective ranges. $C$ can be considered as a random noise added to the flux magnitude to emulate additional fluctuations and nonlinearities observed in practice in the actual flux profile due to changes in contact and interfacial heat transfer between mold and steel \cite{zappulla2020multiphysics,koric2006efficient}. After all control points and the flux values are defined, a radial basis interpolation with a Gaussian function is used to interpolate the flux. A total of 4000 FE simulations were generated with distinct heat flux histories, and a typical example of the time-dependent heat flux is shown in \fref{heat_flux}.

\subsubsection{History-dependent plastic deformation}
\label{sec:plastic_def}
\begin{figure}[b!] 
    \centering
         \includegraphics[width=0.6\textwidth]{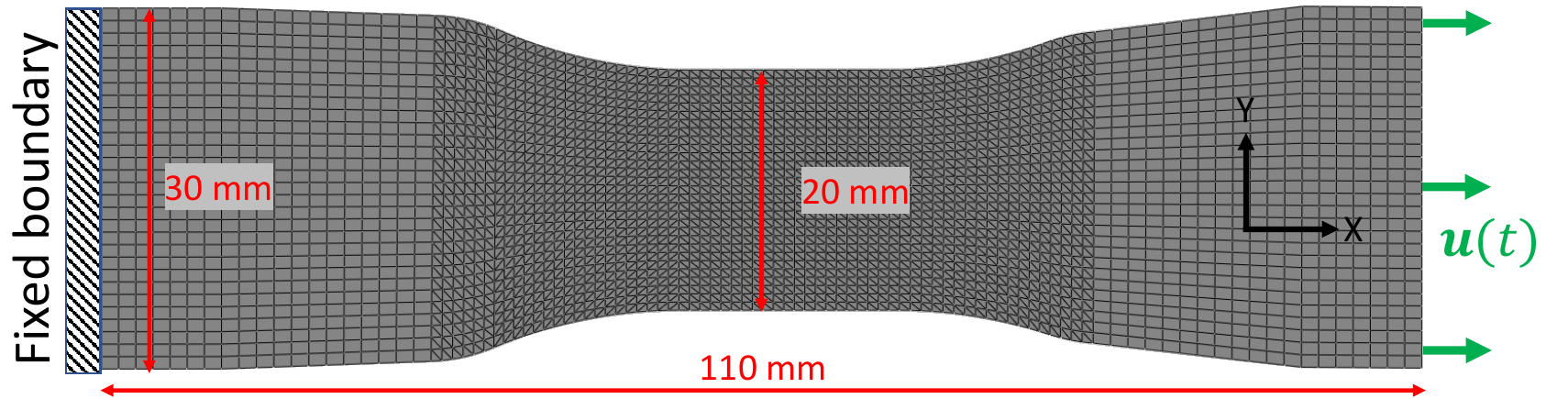}
    \caption{Schematic of the dogbone specimen and the mesh used in FE simulations. The applied displacement is along the global X axis.}
    \label{dogbone_domain}
\end{figure}

In the second example, plastic deformation of a dog bone specimen under a time-dependent loading history is studied, which follows from the recent work by Koric et al. \cite{koric2023deep}. The dog bone specimen has a length of 110 mm and a width (at the grip section) of 30 mm, with a gauge region width of 20 mm. A schematic of the domain along with the mesh used in FE simulations are shown in \fref{dogbone_domain}.
A total of 4756 linear plane stress elements (four-node quadrilaterals and three-node triangles) were used to mesh the specimen with a plane-stress thickness of 1 mm. In the absence of any body and inertial forces, the equilibrium equations and boundary conditions can be stated in terms of the Cauchy stress $\bm{\sigma}$ as:
\begin{equation}
\begin{aligned}
    \nabla \cdot \bm{\sigma} = \bm{0}, \;\; \forall \bm{X} \in \Omega,\\
    \bm{ u } = \Bar{\bm{u}}, \;\; \forall \bm{X} \in \partial \Omega_u,\\
    \bm{\sigma} \cdot \bm{ n } = \Bar{\bm{t}} , \;\; \forall \bm{X} \in \partial \Omega_t,
    \label{strong}
\end{aligned}
\end{equation}
where $\bm{n}$, $\Bar{\bm{u}}$, and $\Bar{\bm{t}}$ denote the outward boundary normal, prescribed displacement, and prescribed traction, respectively. Under small deformation assumption, total strain is given by:
\begin{equation}
    \bm{\epsilon} = \frac{1}{2} ( \nabla \bm{u} + \nabla \bm{u}^T ),
    \label{strain}
\end{equation}
The small-strain formulation of plasticity was used, so the total strain is decomposed additively into its elastic and plastic parts:
\begin{equation}
    \bm{\epsilon} = \bm{\epsilon}^e + \bm{\epsilon}^p.
    \label{decomposition}
\end{equation}
For linear elastic and isotropic material under plane stress condition, the constitutive equation is:
\begin{equation}
    \begin{bmatrix}
    \sigma_{11} \\
    \sigma_{22} \\
    \sigma_{12}
    \end{bmatrix}
    =
    \begin{bmatrix}
    \frac{E}{1-\nu^2} & \frac{\nu E}{1-\nu^2} & 0 \\
    \frac{\nu E}{1-\nu^2} & \frac{E}{1-\nu^2} & 0 \\
    0 & 0 & \frac{E}{2(1+\nu)}
    \end{bmatrix}
    \begin{bmatrix}
    \epsilon_{11} \\
    \epsilon_{22} \\
    \epsilon_{12}
    \end{bmatrix},
    \label{stress}
\end{equation}
where $E$ and $\nu$ are the Young's modulus and Poisson's ratio. In this work, $J_2$ plasticity with linear isotropic hardening was used:
\begin{equation}
    \sigma_y( \Bar{\epsilon}_p ) = \sigma_{y0} + H \Bar{\epsilon}_p,
    \label{hardening}
\end{equation}
where $\sigma_y$, $\Bar{\epsilon}_p$, $\sigma_{y0}$, and $H$ denote the flow stress, equivalent plastic strain, initial yield stress, and the hardening modulus, respectively. The material properties of the elastic-plastic material response are presented in \tref{mat_props}. The material model was integrated implicitly in Abaqus/Standard \citep{Abaqus2021}. 

\begin{table}[h!]
    \caption{Material properties of the elastic-plastic material model}
    \small
    \centering
    \begin{tabular}{cccccccc}
     Property & \vline & $E$ [MPa] & $\nu$ [/] & $\sigma_{y0}$ [MPa] & H [MPa]\\
    \hline
    Value & \vline  & 2.09$\times 10^{5}$ & 0.3 & 235 & 800\\
    \end{tabular}
    \label{mat_props}
\end{table}

The specimen was fixed on the left side and a prescribed, time-dependent displacement was applied on the right edge. Six control points were used to define the loading path. Besides the two end points at $t=0$ and $t=1s$, four other control points were randomly sampled from the range $[0.1,0.9]s$. The applied displacement is 0 at $t=0$. The displacement magnitude at each control point was randomly selected such that the nominal axial strain magnitude is below 5\%. Radial basis interpolation was used to interpolate the applied displacement at arbitrary time points. A typical example of the applied displacement is shown in \fref{disp}. A total of 15000 FE simulations were generated, and the von Mises stress was stored as the ground truth labels in the NN training and is defined as:
\begin{equation}
    \Bar{\sigma} = \sqrt{ \sigma_{11}^2 + \sigma_{22}^2 + \sigma_{11}\sigma_{22} + 3\sigma_{12}^2 }.
\end{equation}

\section{Results and discussion}
\label{sec:results}
FE simulations of the transient heat transfer and plastic deformation were conducted with eight high-end AMD EPYC 7763 Milan CPU cores. All NN training and inference were conducted using a single Nvidia A100 GPU card on Delta, an HPC cluster hosted at the National Center for Supercomputing Applications (NCSA). To evaluate the model performance in the test set, two quantitative metrics were used. The first one is the relative $L_2$ error, which is given by:
\begin{equation}
    {\rm{Relative \; L_2 \; error}} = \frac{ | f_{FE} - f_{Pred} |_2 }{ |f_{FE}|_2 } \times 100\%,
\end{equation}
where $f_{FE}$, and $f_{Pred}$ denote the FE-simulated field value, and the NN-predicted field value, respectively. The second one is the commonly used $R^2$ value, which is defined as:
\begin{equation}
    R^2 = 1 - \frac{ \sum_{i=1}^{N_T} \left( f_{FE} - f_{Pred} \right)^2 }{  \sum_{i=1}^{N_T} \left( f_{FE} - \Bar{f}_{FE} \right)^2 } ,
\end{equation}
where $N_T$, $\Bar{f}_{FE}$ are the total number of test cases and the mean value of the FE-simulated field values.

\subsection{Transient heat transfer}
\label{sec:ht_res}
\begin{figure}[h!] 
    \centering
     \subfloat[]{
         \includegraphics[trim={0cm 0cm 0cm .2cm},clip,width=0.4\textwidth]{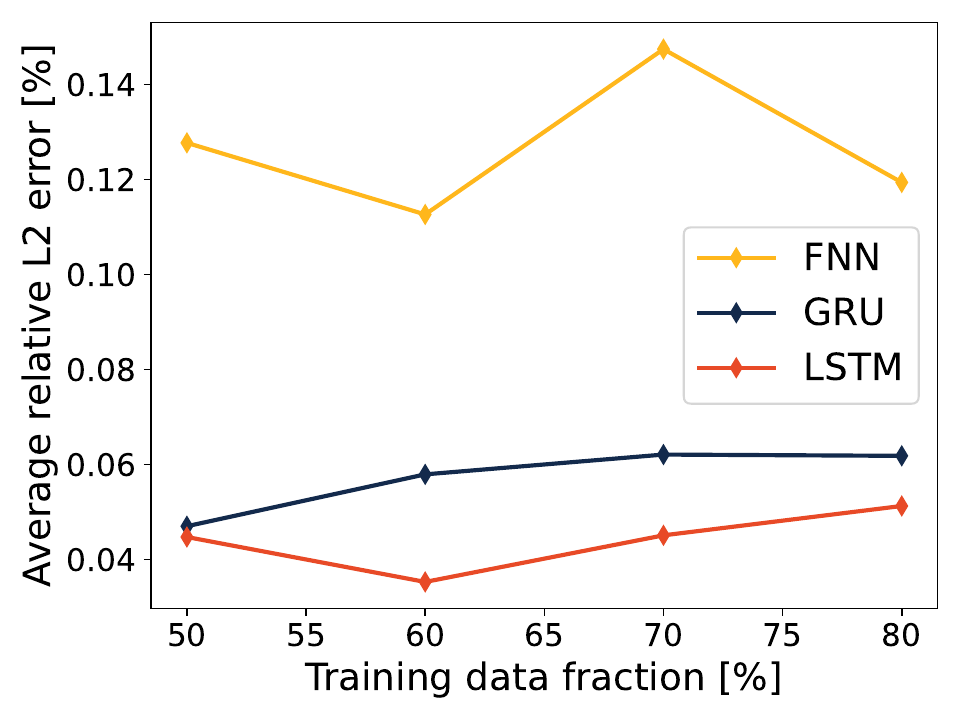}
         \label{htr1}
     }
     \subfloat[]{
         \includegraphics[trim={0cm 0cm 0cm .2cm},clip,width=0.4\textwidth]{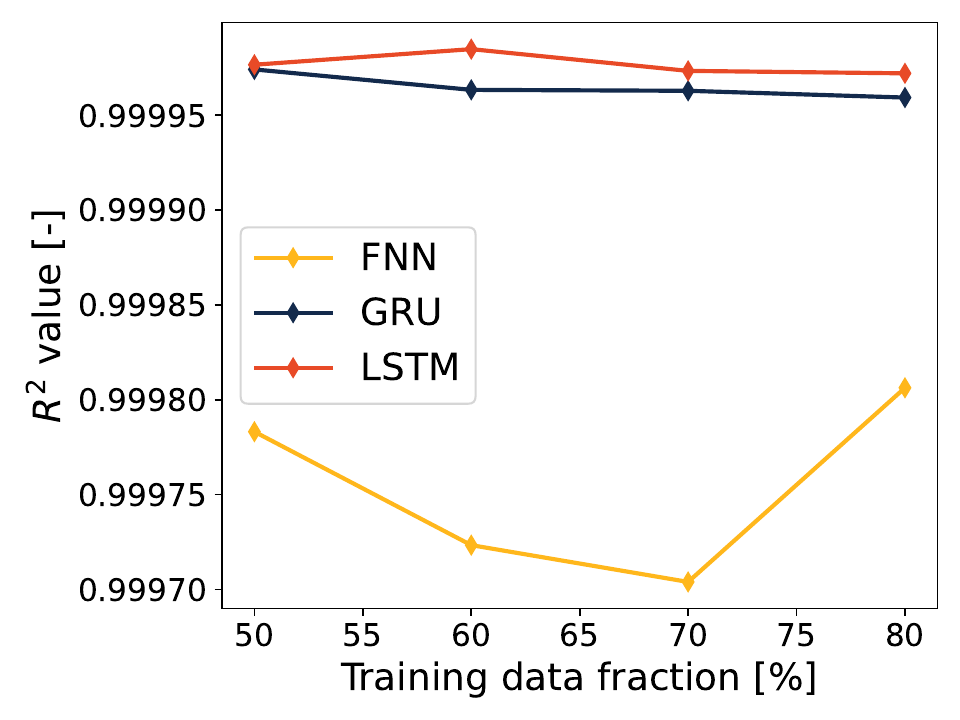}
         \label{htr2}
     }
    \caption{Performance metrics for the three models trained with a different number of training data points.}
    \label{dat_sensiticity1}
\end{figure}

First, the three DeepONet models were trained using different fractions of the available data to investigate their sensitivity of prediction accuracy on the amount of training data. A total of four fractions were studied, they were 50\%, 60\%, 70\%, and 80\%. The results were summarized in \fref{dat_sensiticity1}.

The classical 80-20 split was adopted in subsequent discussions, meaning that 80\% of the data was used in training the models. 5-fold cross-validation was conducted on the best-performing model in \fref{dat_sensiticity1} (i.e., the LSTM-based DeepONet) to test the repeatability of the model performance. The results are summarized in \tref{cv1}.
\begin{table}[h!]
    \caption{5-fold cross-validation, transient heat transfer}
    \small
    \centering
    \begin{tabular}{cccc}
     Model & \vline & Relative $L_2$ error [\%] & $R^2$ value\\
    \hline
    LSTM & \vline  & 6.323$\times 10^{-2}$ (1.383$\times 10^{-2}$) \tablefootnote{Data format: Mean (Standard deviation)} & 1.000 (1.860$\times 10^{-5}$) \\
    \end{tabular}
    \label{cv1}
\end{table}

We compared the LSTM model with the median performance from the 5-fold cross-validation with the FNN- and GRU-DeepONets trained using an 80-20 data split (shown in \fref{dat_sensiticity1}). For the heat transfer problem, training of the FNN-, GRU- and LSTM-based DeepONets took 7210s, 18086s, and 19722s, respectively. The inference time for the three NN models compared to FE simulation time is shown in \tref{solidification_model_times}.
\begin{table}[h!]
    \caption{Computational cost of the heat transfer problem}
    \small
    \centering
    \begin{tabular}{ccccc}
      & \vline & FE simulation time [s] & Inference time [s] & Speed up compared to FE (X)\\
    \hline
    FE simulation & \vline  & 60 & / & / \\
    FNN & \vline  & / & 5.35 $\times 10^{-3}$ & 1.1$\times 10^{4}$ \\
    GRU & \vline  & / & 2.38 $\times 10^{-2}$ & 2.5$\times 10^{3}$ \\
    LSTM & \vline  & / & 3.08 $\times 10^{-2}$ & 1.9$\times 10^{3}$ \\
    \end{tabular}
    \label{solidification_model_times}
\end{table}

\begin{figure}[h!] 
    \centering
     \subfloat[]{
         \includegraphics[trim={0cm 0cm 0cm 0.25cm},clip,width=0.3\textwidth]{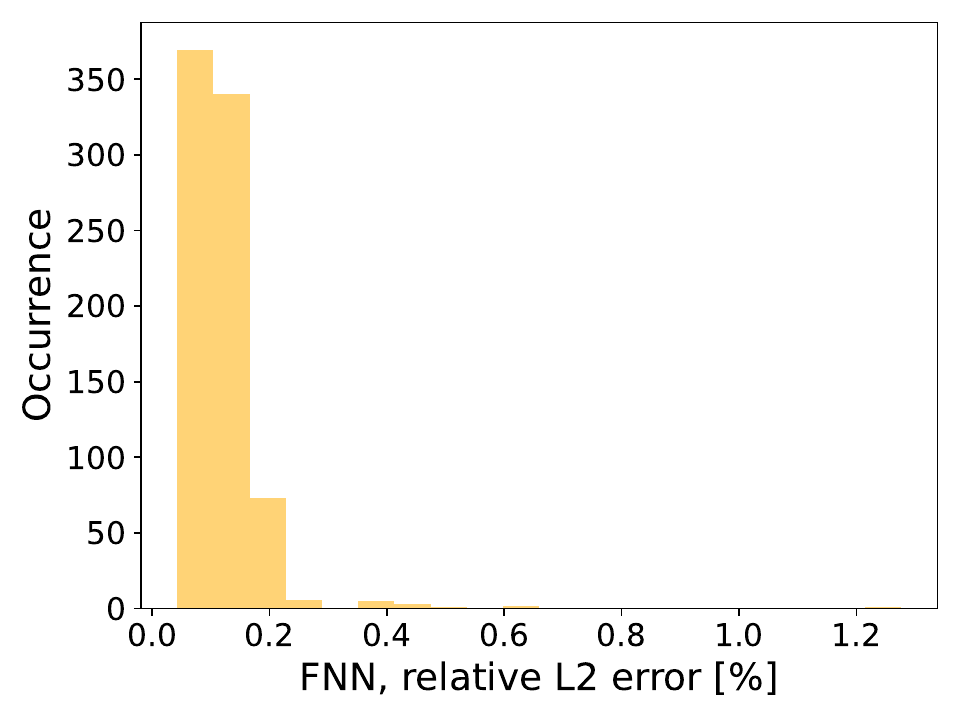}
         \label{er0}
     }
     \subfloat[]{
         \includegraphics[trim={0cm 0cm 0cm 0.25cm},clip,width=0.3\textwidth]{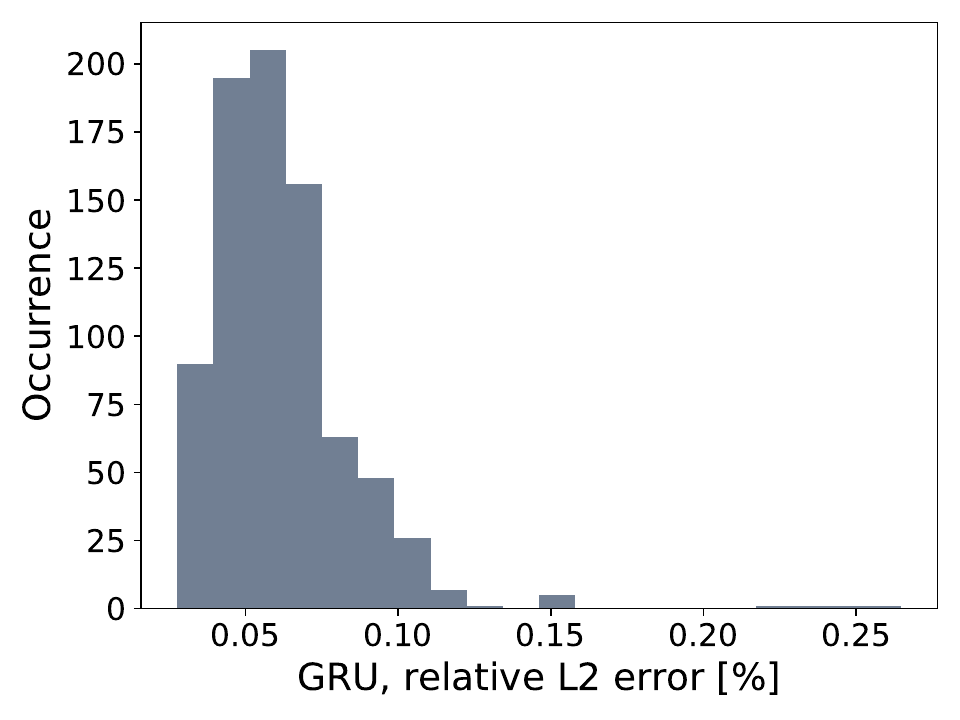}
         \label{er1}
     }
     \subfloat[]{
         \includegraphics[trim={0cm 0cm 0cm 0.25cm},clip,width=0.3\textwidth]{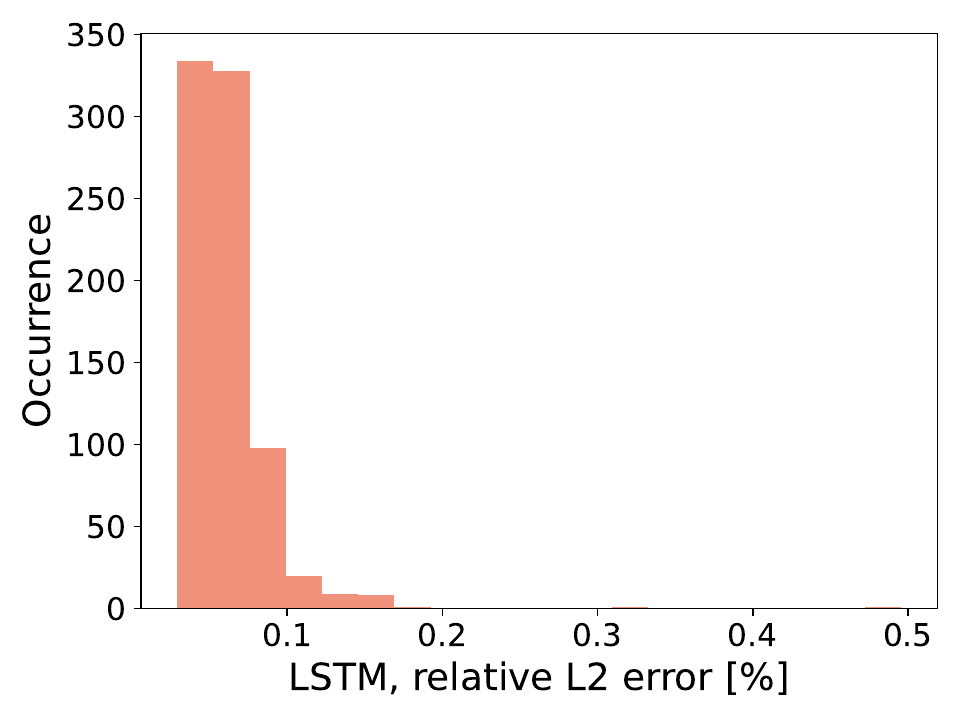}
         \label{er2}
     }
    \caption{Histograms showing the relative $L_2$ error distributions over all test cases for the transient heat transfer problem.}
    \label{solidification_err_hist}
\end{figure}

\begin{figure}[h!]
\newcommand\x{0.3}
    \centering
    \begin{tabular}{ c c c c c }
    \begin{minipage}[c]{\x\textwidth}
       \centering 
        \subfloat[FNN, best]{\includegraphics[trim={1.cm 1.5cm 1.5cm 1.8cm},clip,width=\textwidth]{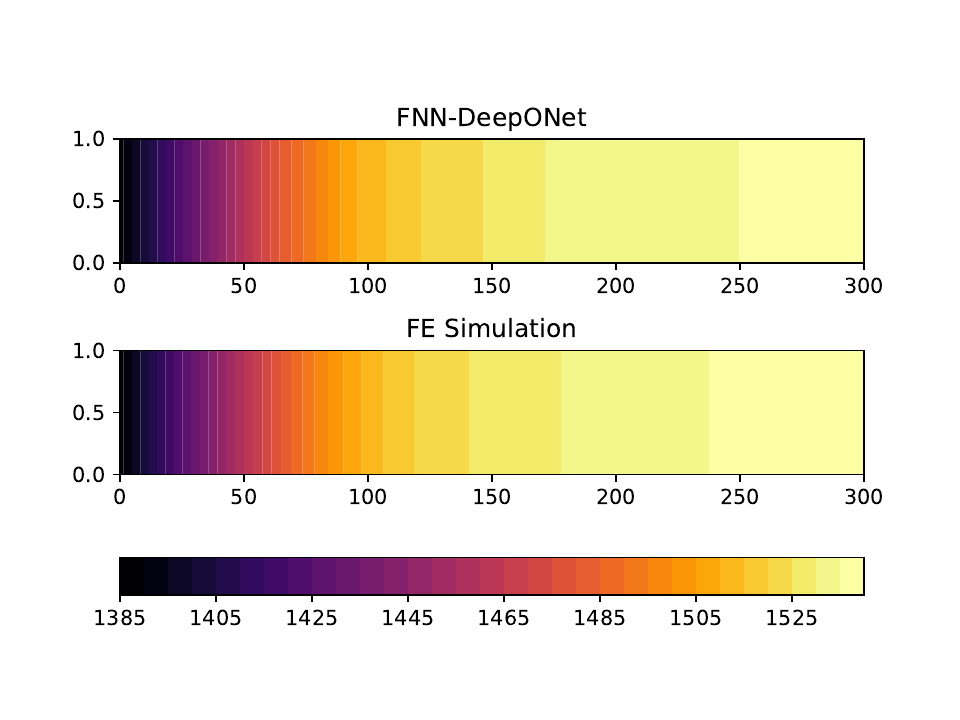}
        \label{ht11}}
    \end{minipage} &
    \begin{minipage}[c]{\x\textwidth}
       \centering 
        \subfloat[FNN, 90$^{th}$ pct]{\includegraphics[trim={1.cm 1.5cm 1.5cm 1.8cm},clip,width=\textwidth]{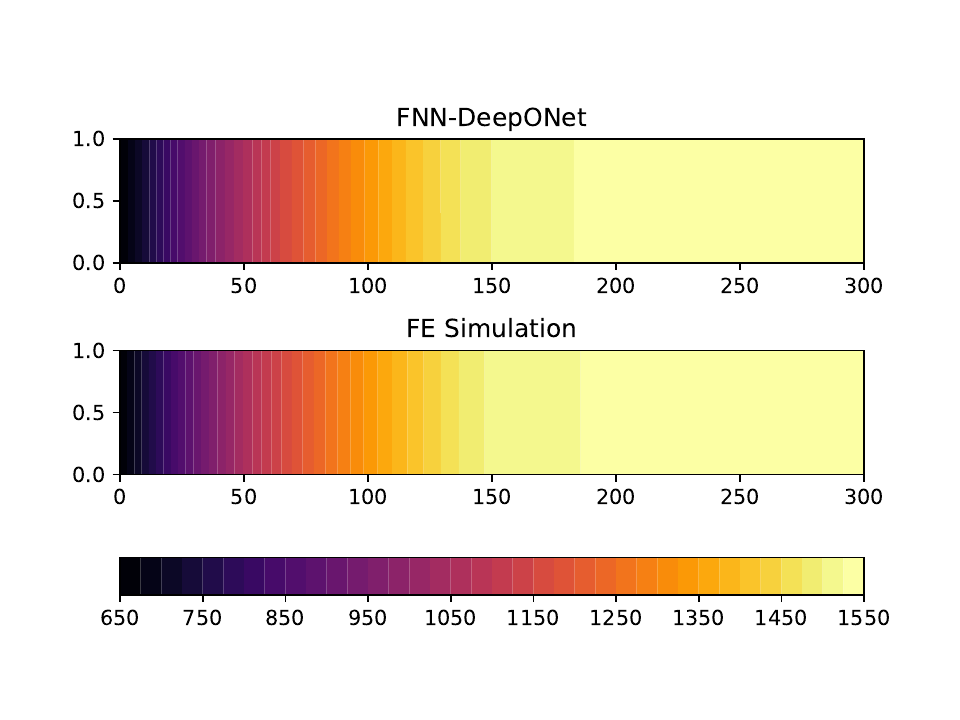}
        \label{ht12}}
    \end{minipage} &
    \begin{minipage}[c]{\x\textwidth}
       \centering 
        \subfloat[FNN, worst]{\includegraphics[trim={1.cm 1.5cm 1.5cm 1.8cm},clip,width=\textwidth]{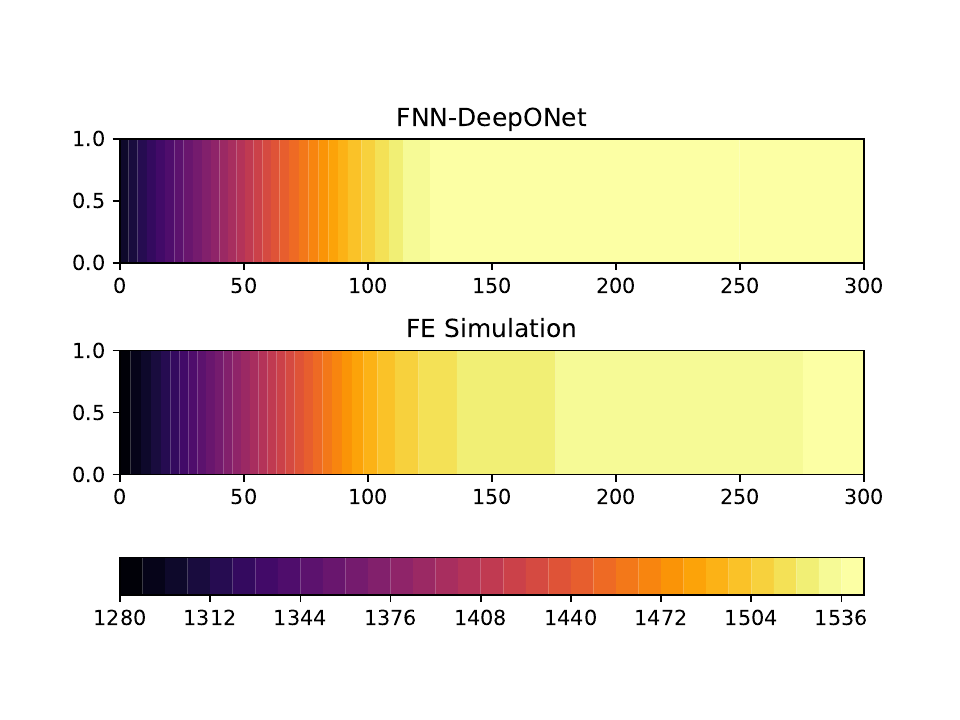}
        \label{ht14}}
    \end{minipage} \\

    \begin{minipage}[c]{\x\textwidth}
       \centering 
        \subfloat[GRU, best]{\includegraphics[trim={1.cm 1.5cm 1.5cm 1.8cm},clip,width=\textwidth]{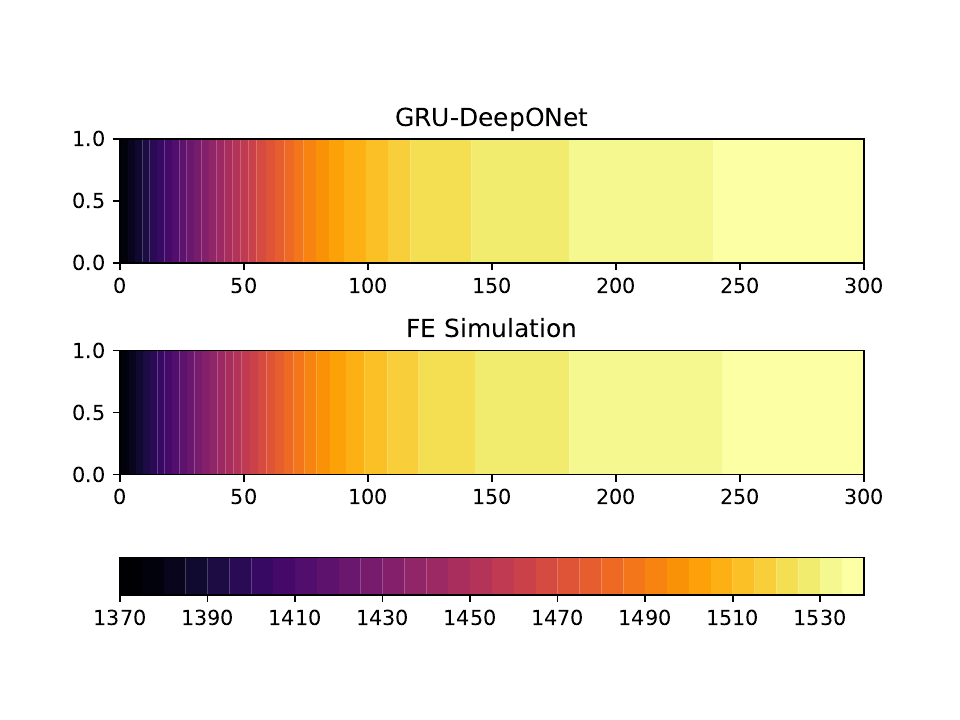}
        \label{ht21}}
    \end{minipage} &
    \begin{minipage}[c]{\x\textwidth}
       \centering 
        \subfloat[GRU, 90$^{th}$ pct]{\includegraphics[trim={1.cm 1.5cm 1.5cm 1.8cm},clip,width=\textwidth]{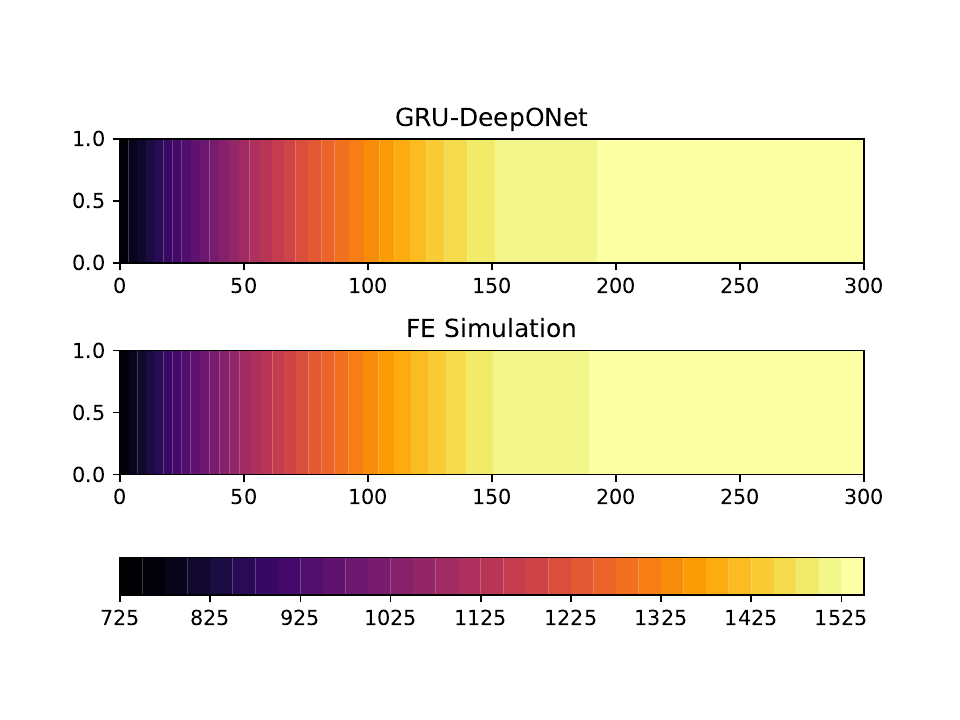}
        \label{ht22}}
    \end{minipage} &
    \begin{minipage}[c]{\x\textwidth}
       \centering 
        \subfloat[GRU, worst]{\includegraphics[trim={1.cm 1.5cm 1.5cm 1.8cm},clip,width=\textwidth]{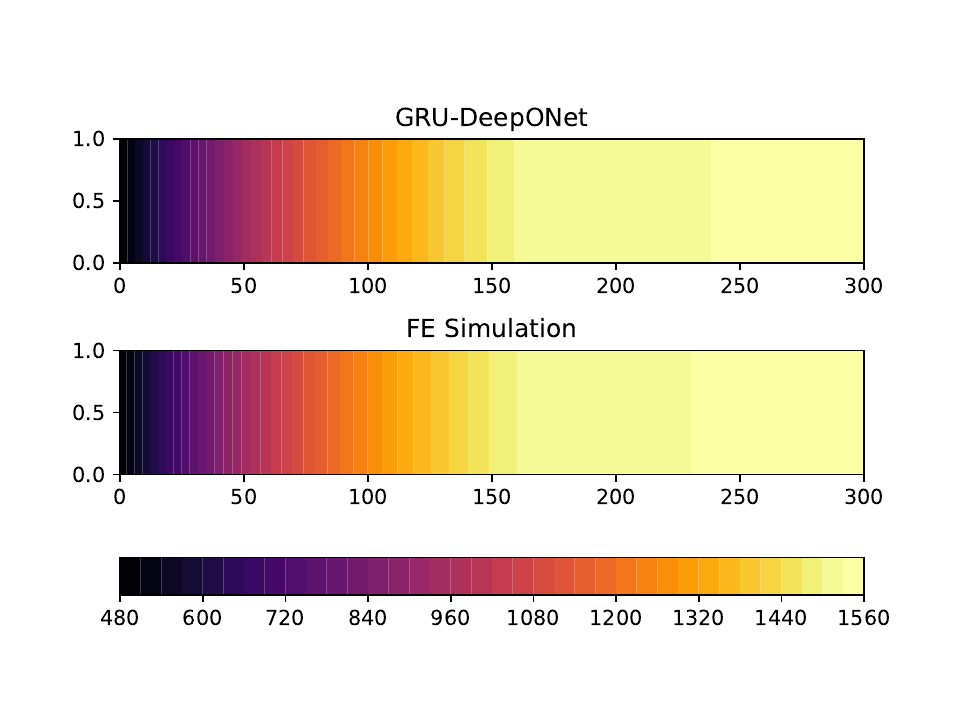}
        \label{ht24}}
    \end{minipage} \\

    \begin{minipage}[c]{\x\textwidth}
       \centering 
        \subfloat[LSTM, best]{\includegraphics[trim={1.cm 1.5cm 1.5cm 1.8cm},clip,width=\textwidth]{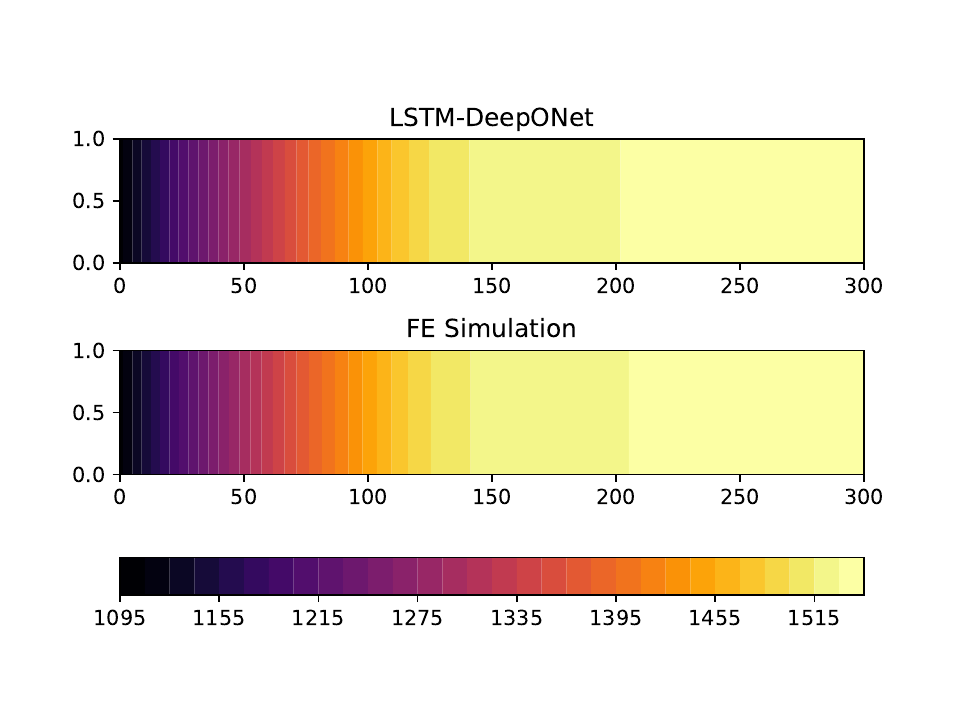}
        \label{ht31}}
    \end{minipage} &
    \begin{minipage}[c]{\x\textwidth}
       \centering 
        \subfloat[LSTM, 90$^{th}$ pct]{\includegraphics[trim={1.cm 1.5cm 1.5cm 1.8cm},clip,width=\textwidth]{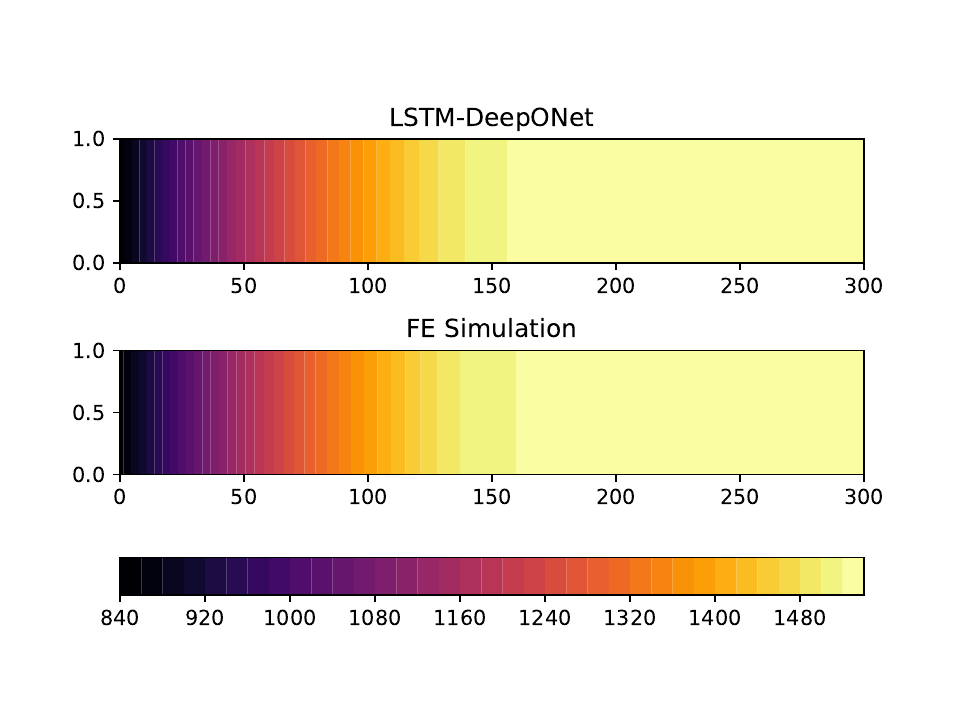}
        \label{ht32}}
    \end{minipage} &
    \begin{minipage}[c]{\x\textwidth}
       \centering 
        \subfloat[LSTM, worst]{\includegraphics[trim={1.cm 1.5cm 1.5cm 1.8cm},clip,width=\textwidth]{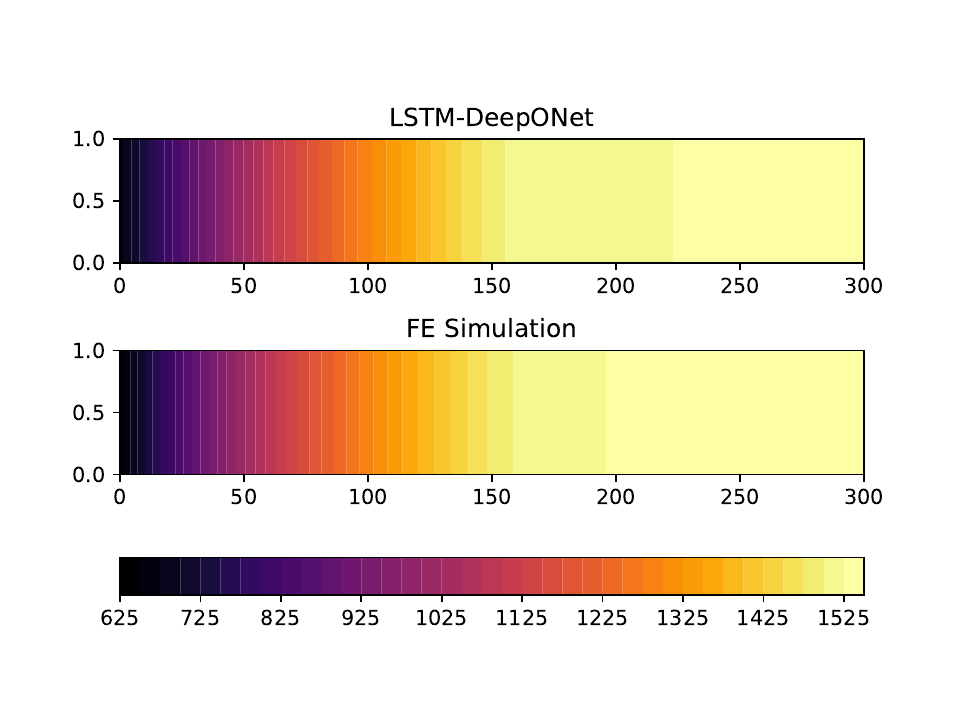}
        \label{ht34}}
    \end{minipage} 
    \end{tabular}
    \caption{Contour plots for the temperature distribution for different DeepONet models at different percentiles of prediction accuracy. }
    \label{solidification_line_plots}
\end{figure}

Key performance statistics on the test set for the three models are shown in \tref{solidification_model_results}. To provide a better perspective of the error distribution over the test cases, histograms of the error distribution are depicted in \fref{solidification_err_hist}. To show the statistical distribution of prediction error among the test cases, final temperature contours that correspond to the 0$^{th}$ (best case), 90$^{th}$ and 100$^{th}$ (worst case) percentile prediction error are displayed in \fref{solidification_line_plots} for all three DeepONet models.
\begin{table}[h!]
    \caption{Error statistics of three DeepONet models on the solidification heat transfer problem}
    \small
    \centering
    \begin{tabular}{ccccc}
      & \vline & Relative $L_2$ error [\%] & Max error [\%] & $R^2$ value \\
    \hline
    FNN & \vline  & 0.119 (0.069) & 1.277 & 0.99981 \\
    GRU & \vline  & 0.062 (0.024) & 0.265 & 0.99996 \\
    LSTM & \vline  & 0.061 (0.028) & 0.496 & 0.99995 \\
    \end{tabular}
    \label{solidification_model_results}
\end{table}

Results from \fref{dat_sensiticity1} indicate that for this simple heat transfer problem, as few as 2000 data points (50\% of all available data) were sufficient to achieve a prediction error of below 0.2\% and $R^2$ value of above 0.999. The performance for the models fluctuates with an increasing number of training data points but remains below 0.2\% for all cases, indicating that overfitting has not occurred. It is also clear from the data that the FNN-based DeepONet consistently under-performs compared to the GRU- and LSTM-based DeepONets, with the LSTM model giving the highest prediction accuracy in this case. However, it is worth noting that the performance difference between GRU- and LSTM-DeepONets is minimal. The results of the 5-fold cross-validation show that the performance of the model is very consistent, with minimal fold-to-fold variation.

The GRU- and LSTM-DeepONet models are more computationally intensive to train than their FNN counterparts, requiring a 1.5 and 1.7 times longer training time, respectively. However, once trained, all three networks can efficiently predict the final stress history with more than 1500 times speedup compared to direct FE simulations. The added training time for the GRU and LSTM models translates to reduced prediction errors, as seen in \tref{solidification_model_results}. In an average sense, across all testing samples, both LSTM- and GRU-based DeepONets were able to lower the $L_2$ error of the FNN-DeepONet by half. However, from \fref{solidification_err_hist}, we see that all three models suffer from outliers with significantly higher prediction errors, which can be as high as 1.3\% for FNN-based DeepONet. The plots in \fref{solidification_line_plots} provide more insight into how the prediction errors are distributed across all testing data points. The GRU- and LSTM-based models can accurately predict the temperature profile even up to the worst-case scenario. However, in the worst-case scenario, the FNN-DeepONet's prediction leads to significant error at the solidification front and the mushy zone, i.e., between solidus and liquidus temperatures. Considering that the thickness of solidifying shell at the
mold exit is calculated from the position of the solidifying front, in this case, the result inferred by classical DeepONet (with FNN in branch) will be of significantly less value for design, optimization, and online controls of this critical steel-making process.

From this example, we see that the performance of the GRU- and LSTM-based DeepONets are highly similar. However, since the GRU-DeepONet has fewer trainable parameters, it trains about 9\% faster than the LSTM model. Therefore, from a computational efficiency and accuracy perspective, it appears that the GRU-DeepONet is the most effective model of the three studied in this work.

\subsection{History-dependent plastic deformation}
\label{sec:plastic_res}
\begin{figure}[h!] 
    \centering
     \subfloat[]{
         \includegraphics[trim={0cm 0cm 0cm .2cm},clip,width=0.4\textwidth]{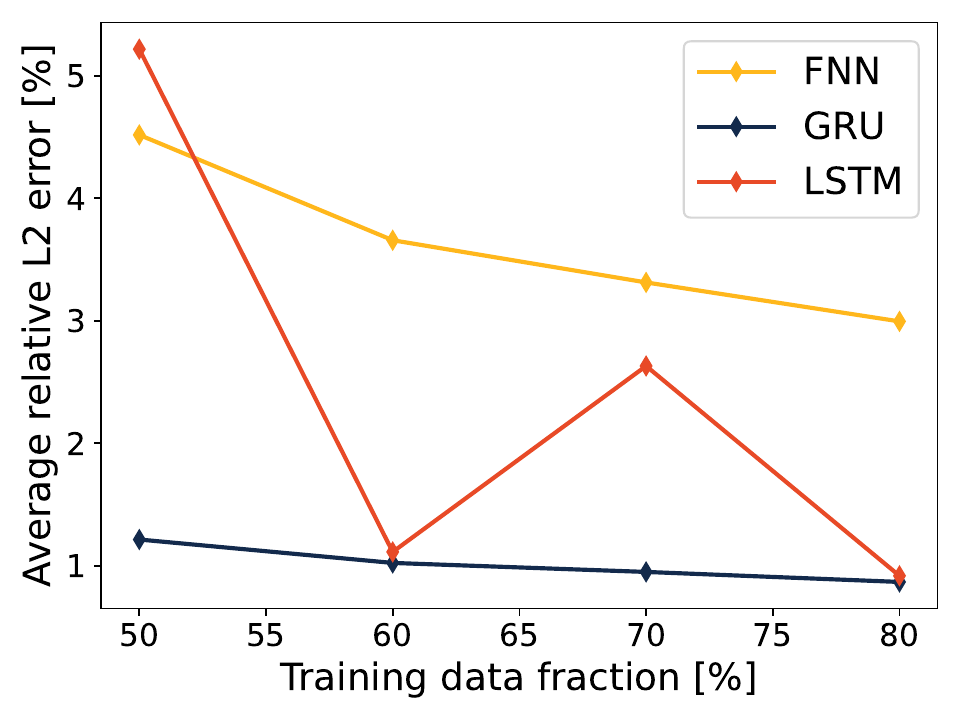}
         \label{htr1}
     }
     \subfloat[]{
         \includegraphics[trim={0cm 0cm 0cm .2cm},clip,width=0.4\textwidth]{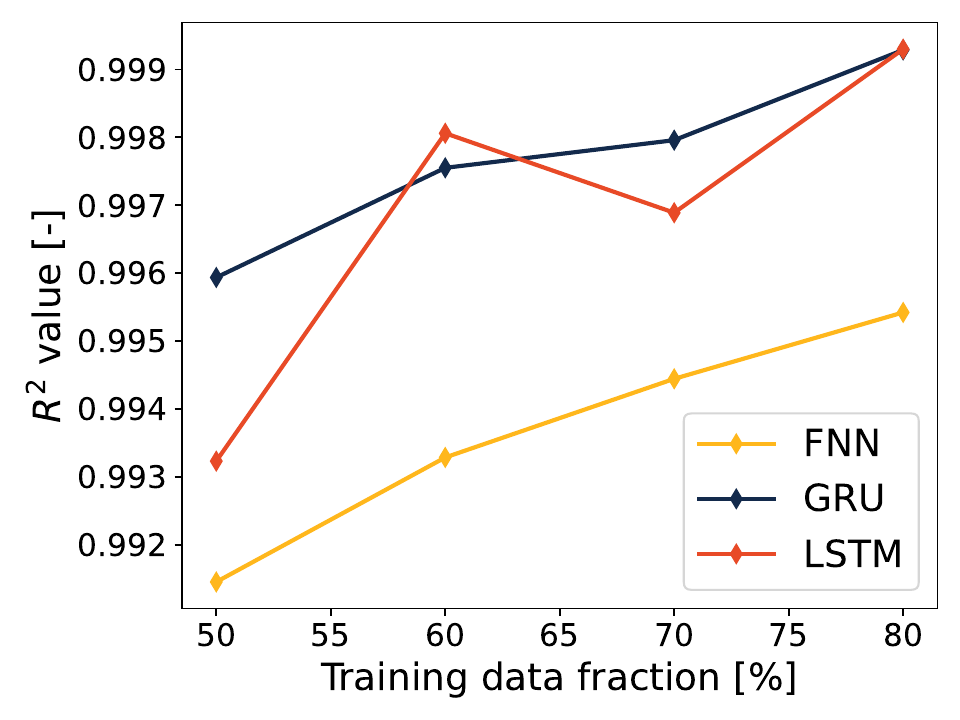}
         \label{htr2}
     }
    \caption{Performance metrics for the three models trained with a different number of training data points.}
    \label{dat_sensiticity2}
\end{figure}

A similar training data fraction study was performed, and the results were summarized in \fref{dat_sensiticity2}.
A 5-fold cross-validation was conducted on the GRU-based DeepONet (best-performing model in \fref{dat_sensiticity2}), and the results are summarized in \tref{cv2}.
\begin{table}[h!]
    \caption{5-fold cross-validation, plastic deformation}
    \small
    \centering
    \begin{tabular}{cccc}
     Model & \vline & Mean relative L2 error [\%] & $R^2$ value\\
    \hline
    GRU & \vline  & 0.902 (0.056) & 0.998 (8.301$\times 10^{-4}$) \\
    \end{tabular}
    \label{cv2}
\end{table}

We compared the GRU model with median performance from the 5-fold cross-validation with the FNN- and LSTM-DeepONets. For the plastic deformation problem, training of the FNN-, GRU- and LSTM-based DeepONets took 7441s, 18145s, and 20730s, respectively. The inference time for the three NN models compared to the FE simulation time is shown in \tref{plasticity_model_times}.
\begin{table}[h!]
    \caption{Computational cost of the plastic deformation problem}
    \small
    \centering
    \begin{tabular}{ccccc}
      & \vline & FE simulation time [s] & Inference time [s] & Speed up compared to FE (X)\\
    \hline
    FE simulation & \vline  & 21 & / & / \\
    FNN & \vline  & / & 3.29 $\times 10^{-2}$ & 6.4$\times 10^{2}$ \\
    GRU & \vline  & / & 8.00 $\times 10^{-2}$ & 2.6$\times 10^{2}$ \\
    LSTM & \vline  & / & 9.70 $\times 10^{-2}$ & 2.2$\times 10^{2}$ \\
    \end{tabular}
    \label{plasticity_model_times}
\end{table}

\begin{figure}[h!] 
    \centering
     \subfloat[]{
         \includegraphics[trim={0cm 0cm 0cm 0.25cm},clip,width=0.3\textwidth]{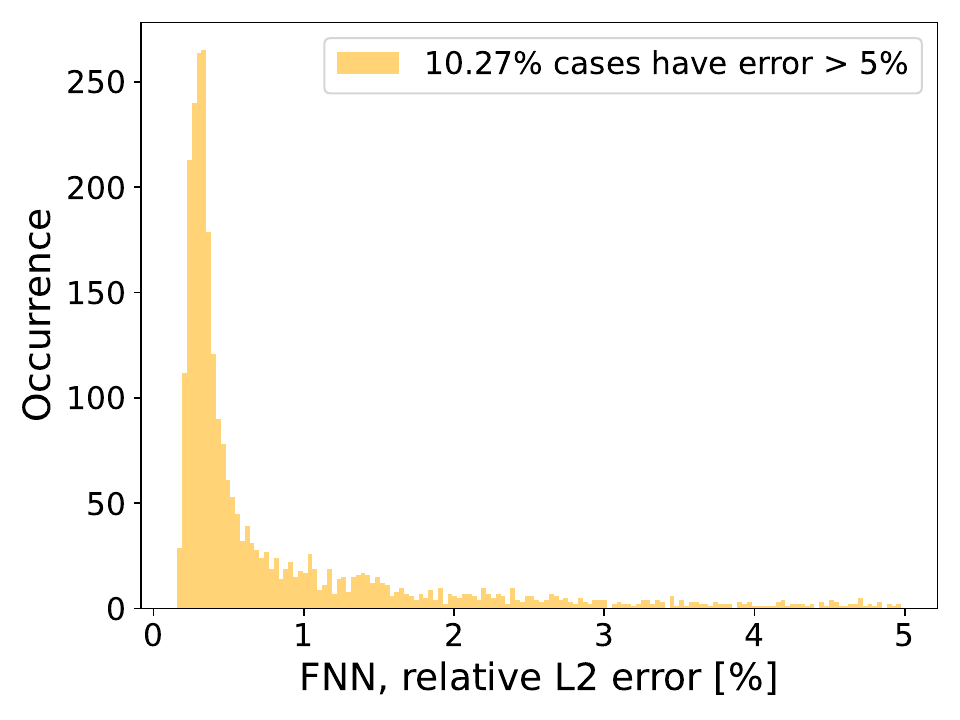}
         \label{erd0}
     }
     \subfloat[]{
         \includegraphics[trim={0cm 0cm 0cm 0.25cm},clip,width=0.3\textwidth]{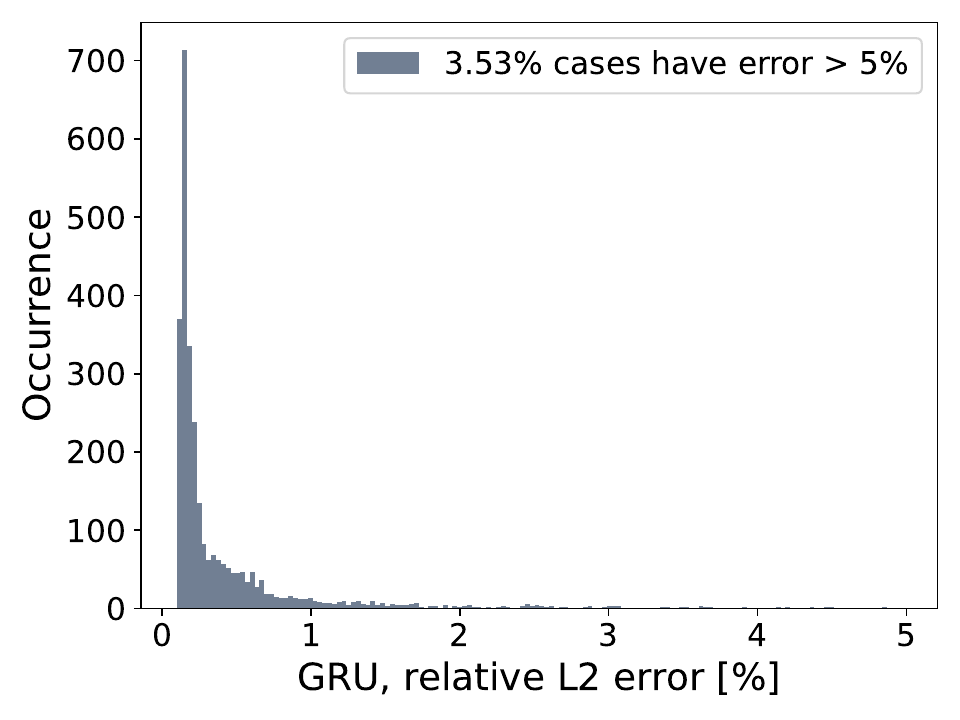}
         \label{erd1}
     }
     \subfloat[]{
         \includegraphics[trim={0cm 0cm 0cm 0.25cm},clip,width=0.3\textwidth]{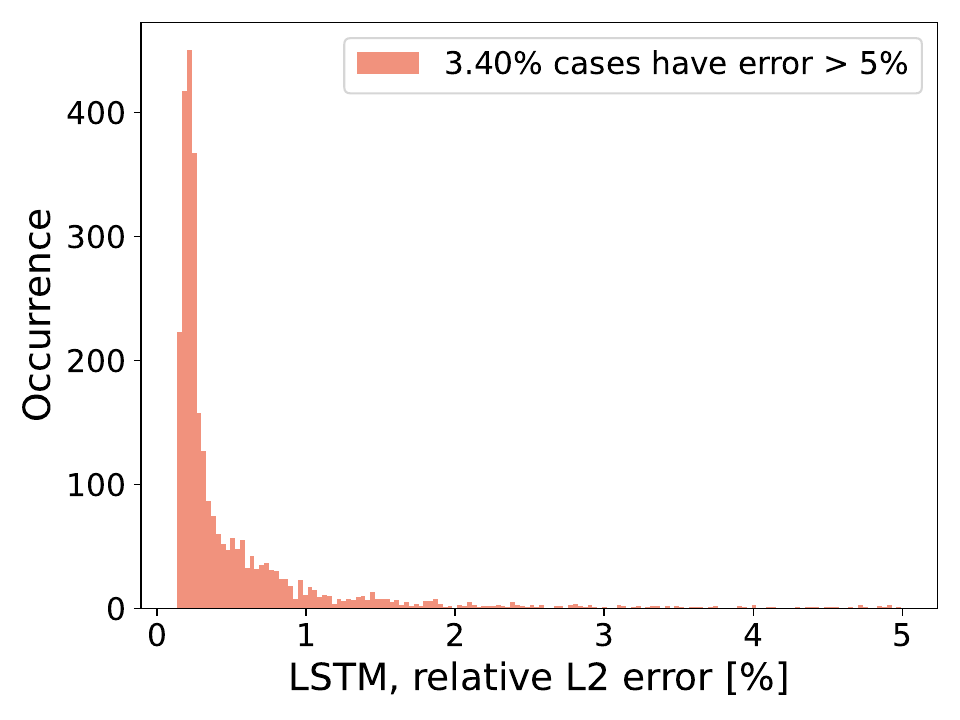}
         \label{erd2}
     }
    \caption{Histograms showing the relative $L_2$ error distributions over all test cases for the plastic deformation of the dog bone specimen. Cases with relative $L_2$ error greater than 5\% were not shown in the histogram and were instead reported on the legend of each plot.}
    \label{plastic_err_hist}
\end{figure}

\begin{figure}[h!]
\newcommand\x{0.29}
    \centering
    \begin{tabular}{ c c c c }
    \begin{minipage}[c]{\x\textwidth}
       \centering 
        \subfloat[FNN, best]{\includegraphics[trim={2cm 1.5cm 2cm 1.2cm},clip,width=\textwidth]{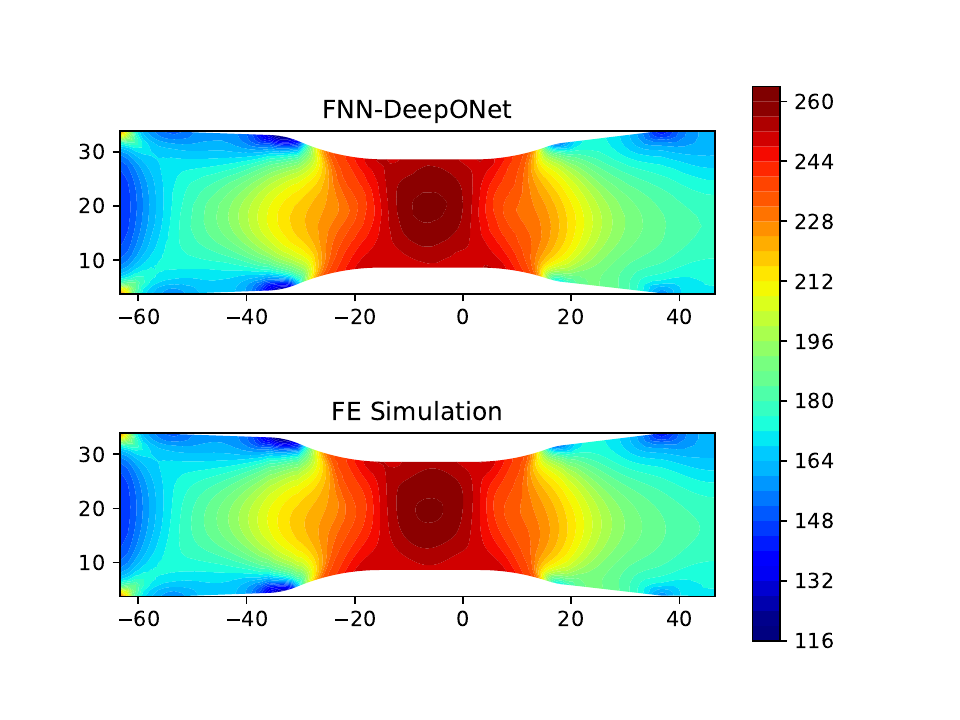}
        \label{pl11}}
    \end{minipage} &
    \begin{minipage}[c]{\x\textwidth}
       \centering 
        \subfloat[FNN, 90$^{th}$ pct]{\includegraphics[trim={2cm 1.5cm 2cm 1.2cm},clip,width=\textwidth]{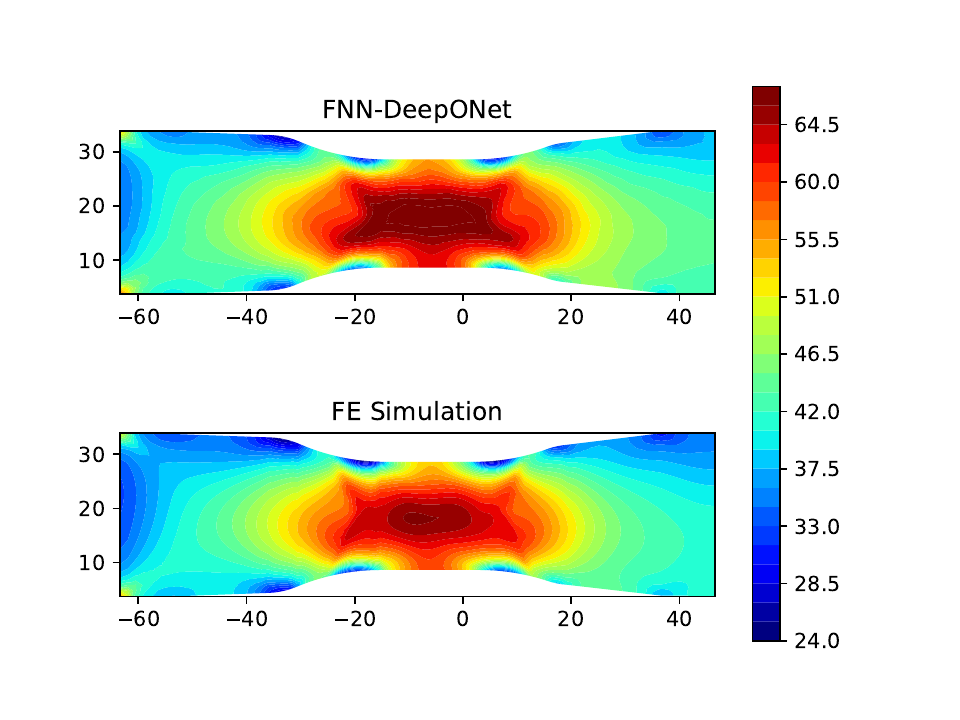}
        \label{pl12}}
    \end{minipage} &
    \begin{minipage}[c]{\x\textwidth}
       \centering 
        \subfloat[FNN, 99$^{th}$ pct]{\includegraphics[trim={2cm 1.5cm 2cm 1.2cm},clip,width=\textwidth]{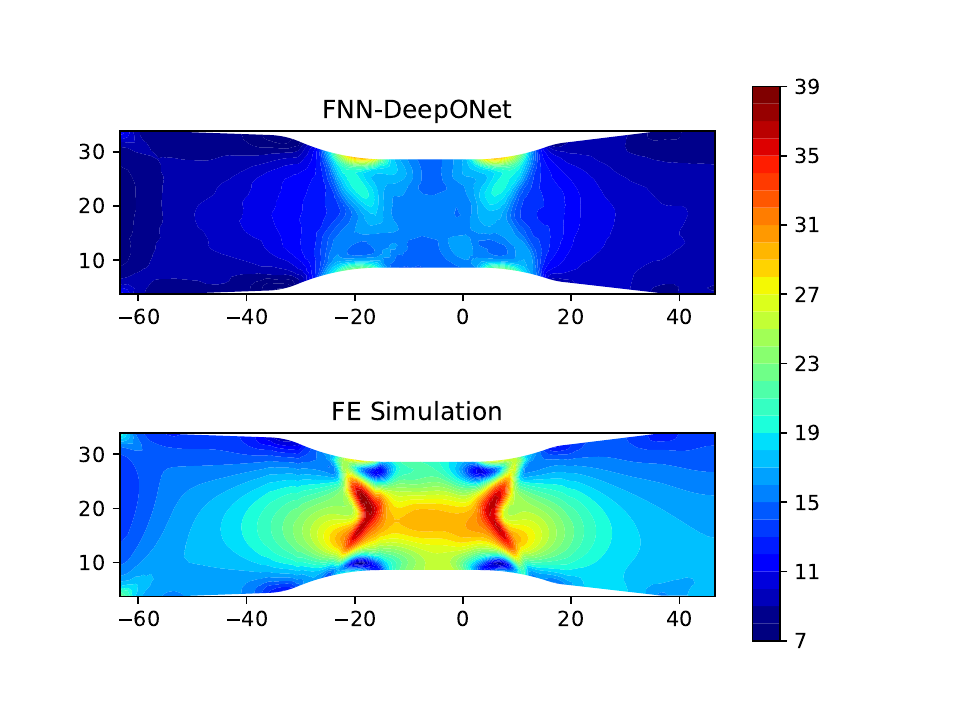}
        \label{pl13}}
    \end{minipage} \\

    \begin{minipage}[c]{\x\textwidth}
       \centering 
        \subfloat[GRU, best]{\includegraphics[trim={2cm 1.5cm 2cm 1.2cm},clip,width=\textwidth]{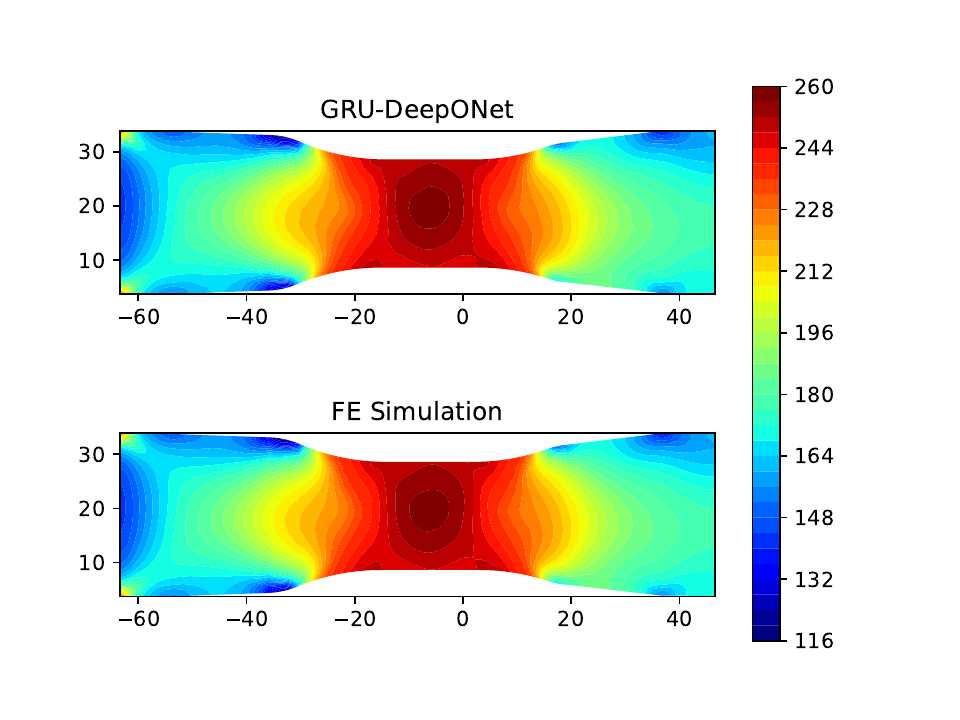}
        \label{pl21}}
    \end{minipage} &
    \begin{minipage}[c]{\x\textwidth}
       \centering 
        \subfloat[GRU, 90$^{th}$ pct]{\includegraphics[trim={2cm 1.5cm 2cm 1.2cm},clip,width=\textwidth]{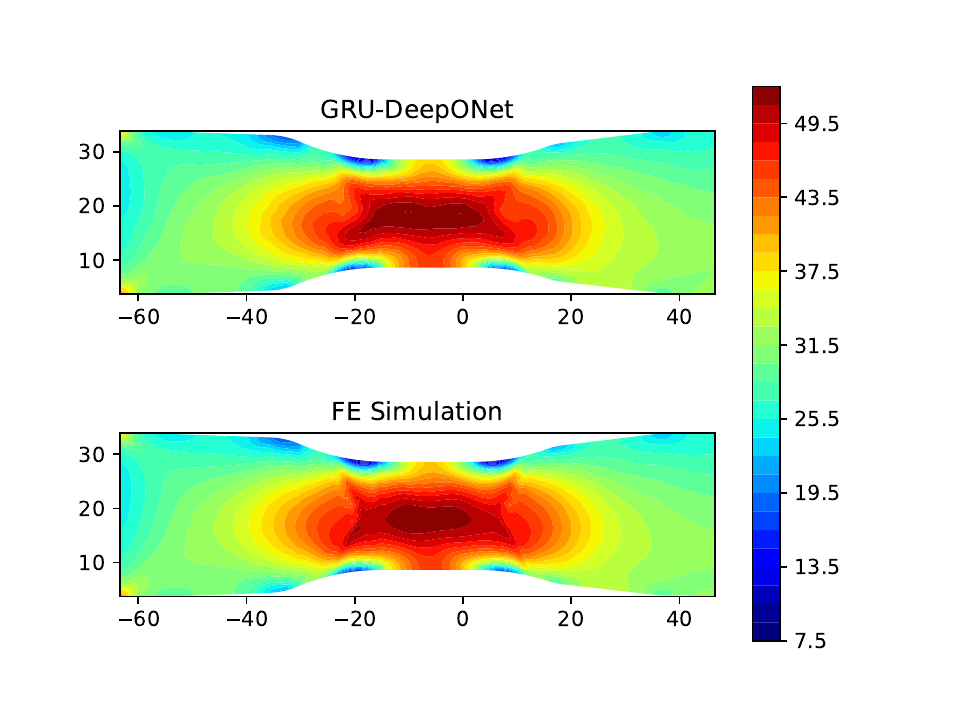}
        \label{pl22}}
    \end{minipage} &
    \begin{minipage}[c]{\x\textwidth}
       \centering 
        \subfloat[GRU, 99$^{th}$ pct]{\includegraphics[trim={2cm 1.5cm 2cm 1.2cm},clip,width=\textwidth]{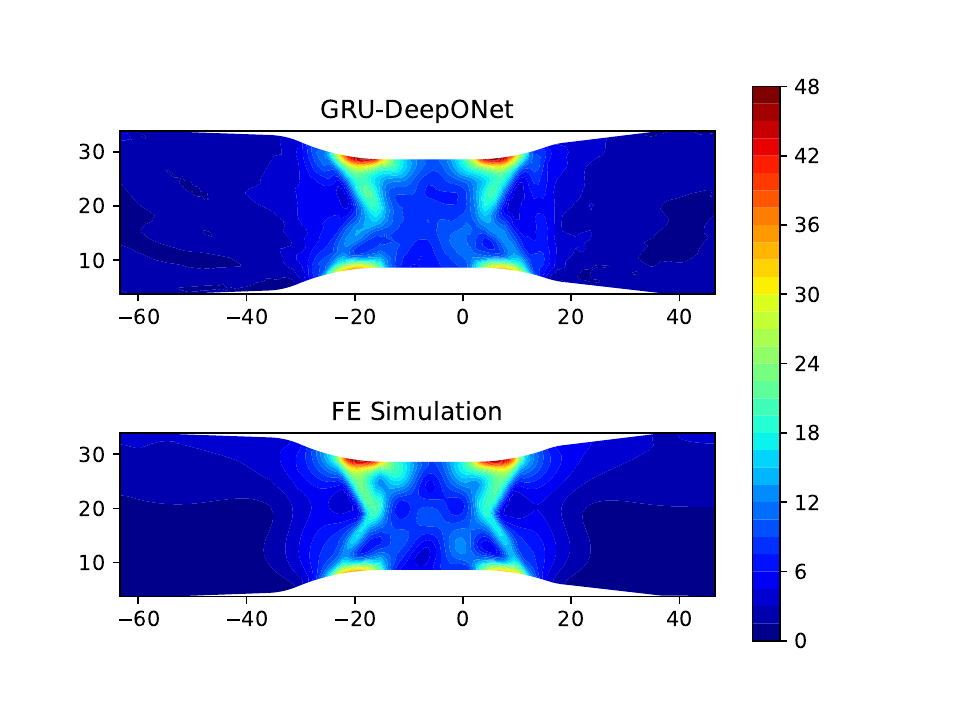}
        \label{pl23}}
    \end{minipage} \\

    \begin{minipage}[c]{\x\textwidth}
       \centering 
        \subfloat[LSTM, best]{\includegraphics[trim={2cm 1.5cm 2cm 1.2cm},clip,width=\textwidth]{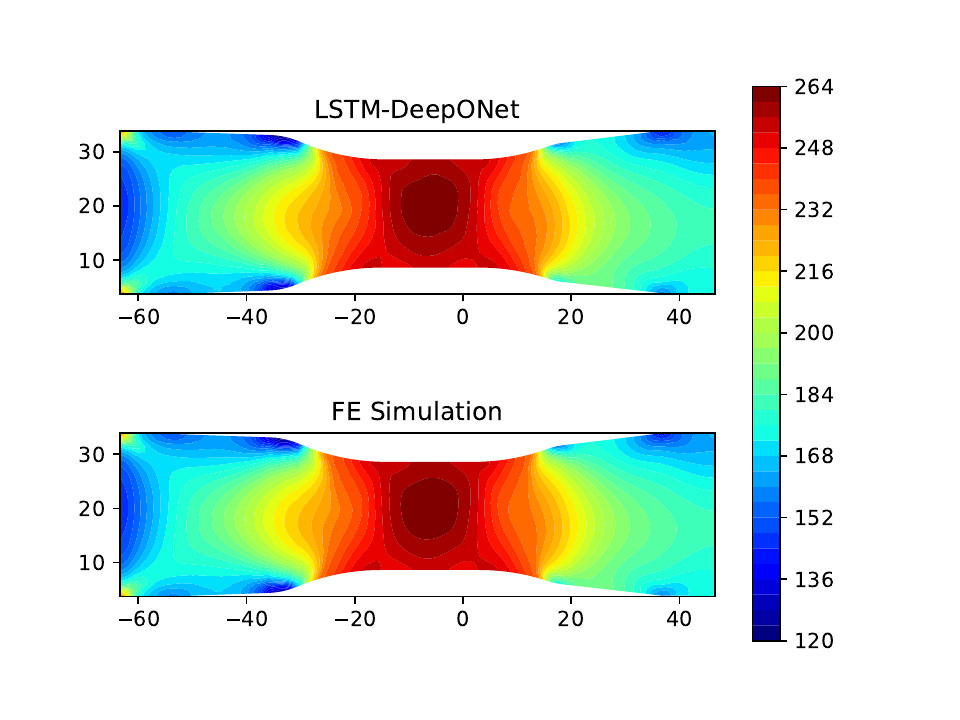}
        \label{pl31}}
    \end{minipage} &
    \begin{minipage}[c]{\x\textwidth}
       \centering 
        \subfloat[LSTM, 90$^{th}$ pct]{\includegraphics[trim={2cm 1.5cm 2cm 1.2cm},clip,width=\textwidth]{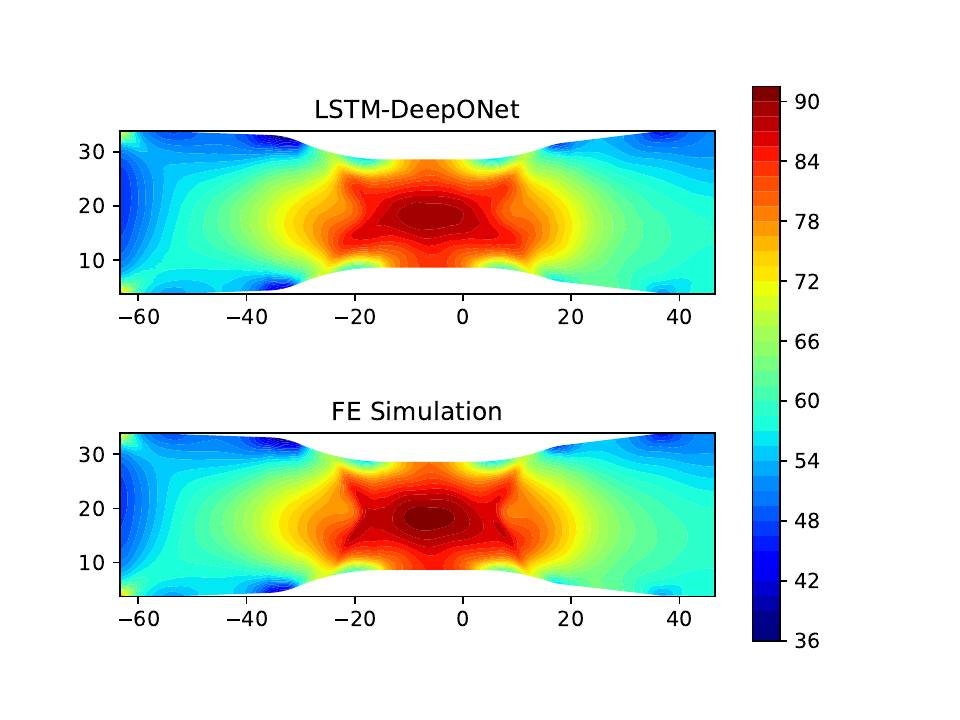}
        \label{pl32}}
    \end{minipage} &
    \begin{minipage}[c]{\x\textwidth}
       \centering 
        \subfloat[LSTM, 99$^{th}$ pct]{\includegraphics[trim={2cm 1.5cm 2cm 1.2cm},clip,width=\textwidth]{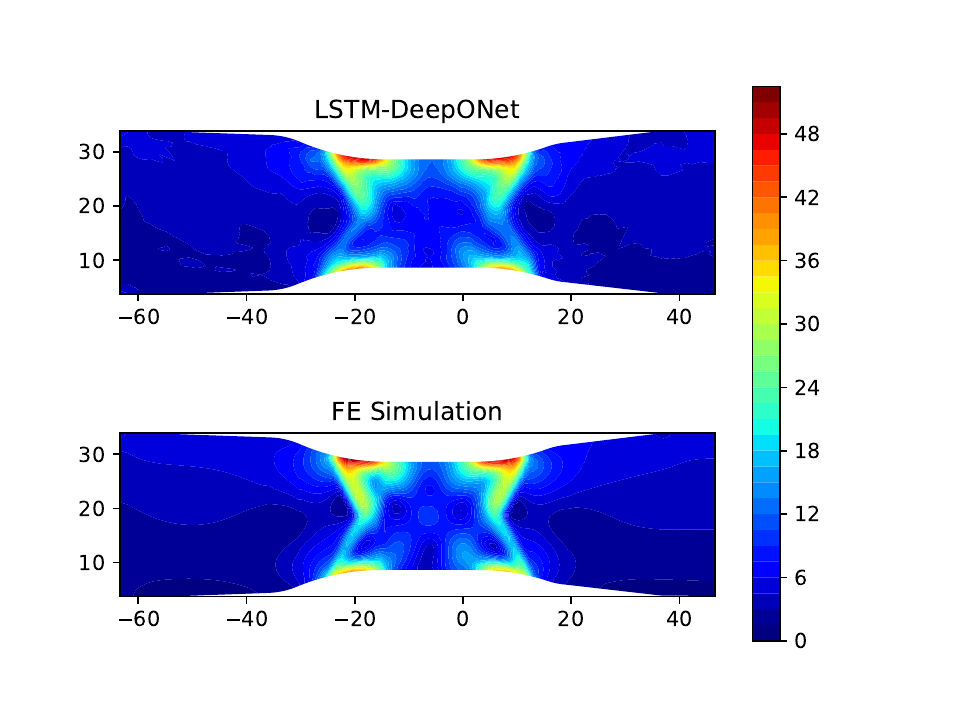}
        \label{pl33}}
    \end{minipage}
    \end{tabular}
    \caption{Contour plots for the Mises stress distribution for different DeepONet models at different percentiles of prediction accuracy. }
    \label{plastic_contour_results}
\end{figure}

Key performance metrics for the three models are shown in \tref{plasticity_model_results}. Histograms of the error distribution are depicted in \fref{plastic_err_hist}. To remove the effect of outliers on the X-axis scaling, the cases with relative $L_2$ error greater than 5\% were not shown in the histograms, and were instead counted and reported in the figure legends. Contour plots of the von Mises stress are displayed in \fref{plastic_contour_results} to show the spatial distribution of prediction errors. Cases correspond to the 0$^{th}$ (best case), 90$^{th}$, and 99$^{th}$ percentile prediction errors for all three DeepONet models.
\begin{table}[h!]
    \caption{Error statistics of three DeepONet models on the plastic deformation problem}
    \small
    \centering
    \begin{tabular}{ccccc}
      & \vline & Relative $L_2$ error [\%] & Max error [\%] & $R^2$ value \\
    \hline
    FNN & \vline  & 2.995 (14.551) & 615.369 & 0.99542 \\
    GRU & \vline  & 0.847 (2.853) & 89.044 & 0.99721 \\
    LSTM & \vline  & 0.919 (2.723) & 70.366 & 0.99930 \\
    \end{tabular}
    \label{plasticity_model_results}
\end{table}

\begin{figure}[h!]
\captionsetup[subfigure]{labelformat=empty}
    \centering
    \begin{tabular}{ c c c }
    \begin{minipage}[c]{.25\textwidth}
       \centering 
        \subfloat[]{\includegraphics[trim={0cm 0cm 0cm 0cm},clip,width=\textwidth]{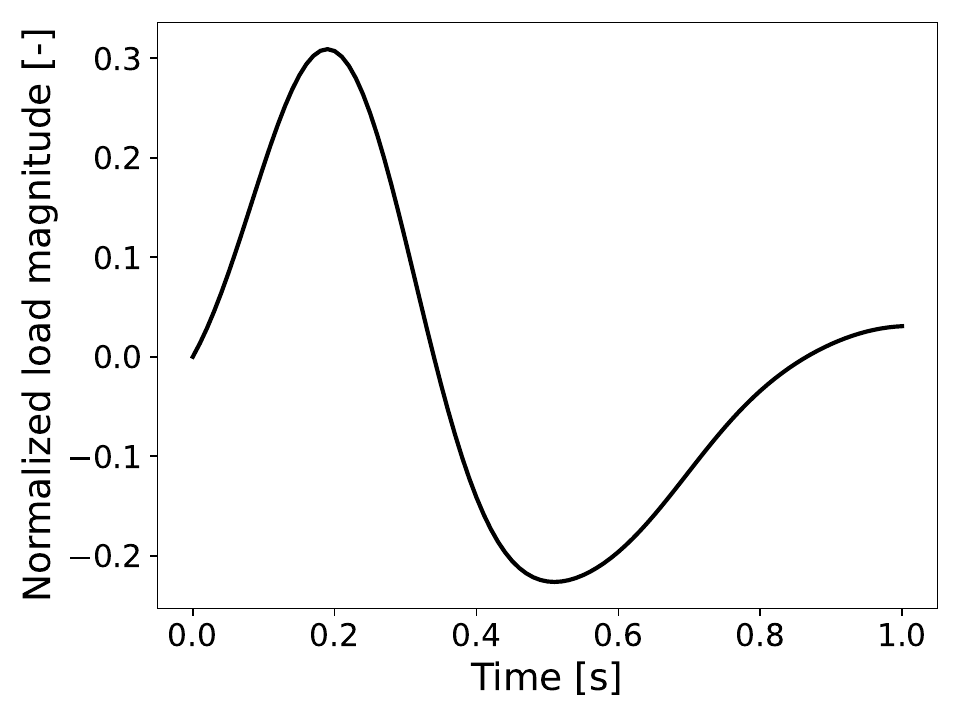}
        \label{fnn_mag}}
    \end{minipage} &
    \begin{minipage}[c]{.5\textwidth}
       \centering 
        \subfloat[(a) FNN, worst case]{\includegraphics[trim={1cm 1.5cm 1cm 5.7cm},clip,width=\textwidth]{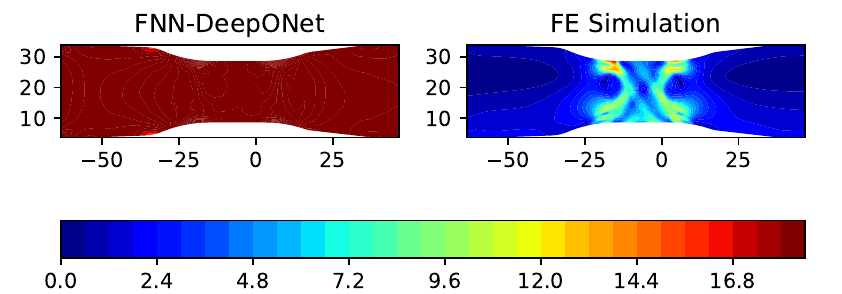}
        \label{fnn_worst}}
    \end{minipage} \\

    \begin{minipage}[c]{.25\textwidth}
       \centering 
        \subfloat[]{\includegraphics[trim={0cm 0cm 0cm 0cm},clip,width=\textwidth]{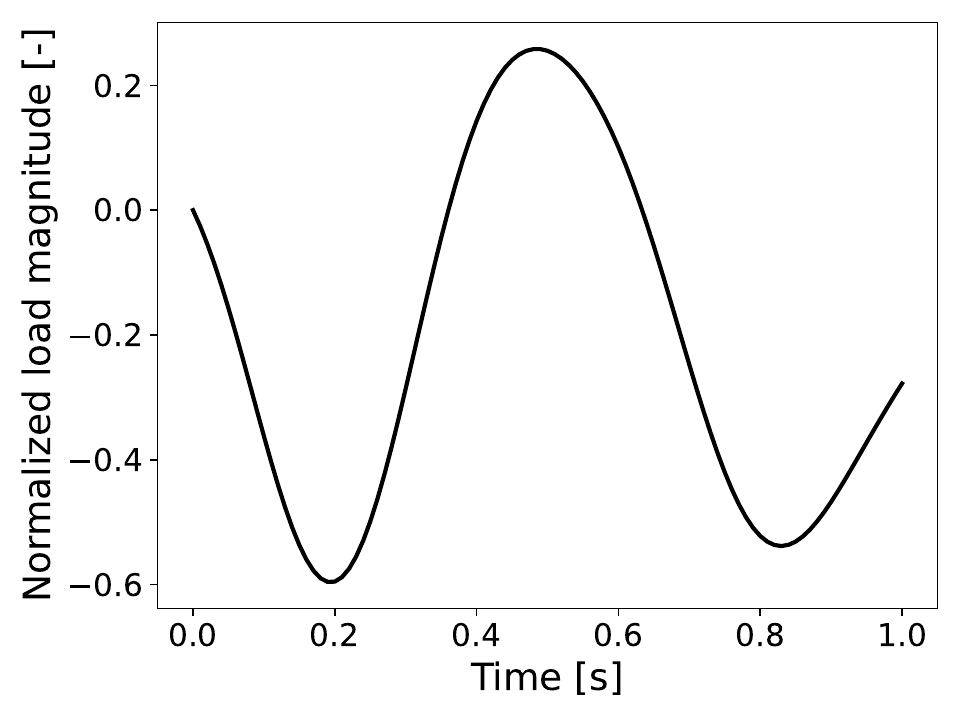}
        \label{gru_mag}}
    \end{minipage} &
    \begin{minipage}[c]{.5\textwidth}
       \centering 
        \subfloat[(b) GRU, worst case]{\includegraphics[trim={1cm 1.5cm 1cm 5.7cm},clip,width=\textwidth]{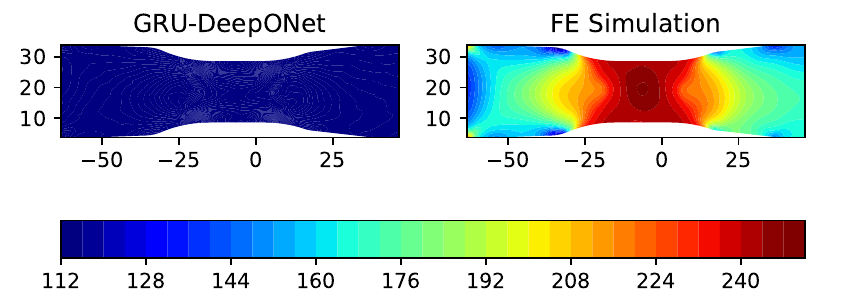}
        \label{gru_worst}}
    \end{minipage} \\

    \begin{minipage}[c]{.25\textwidth}
       \centering 
        \subfloat[]{\includegraphics[trim={0cm 0cm 0cm 0cm},clip,width=\textwidth]{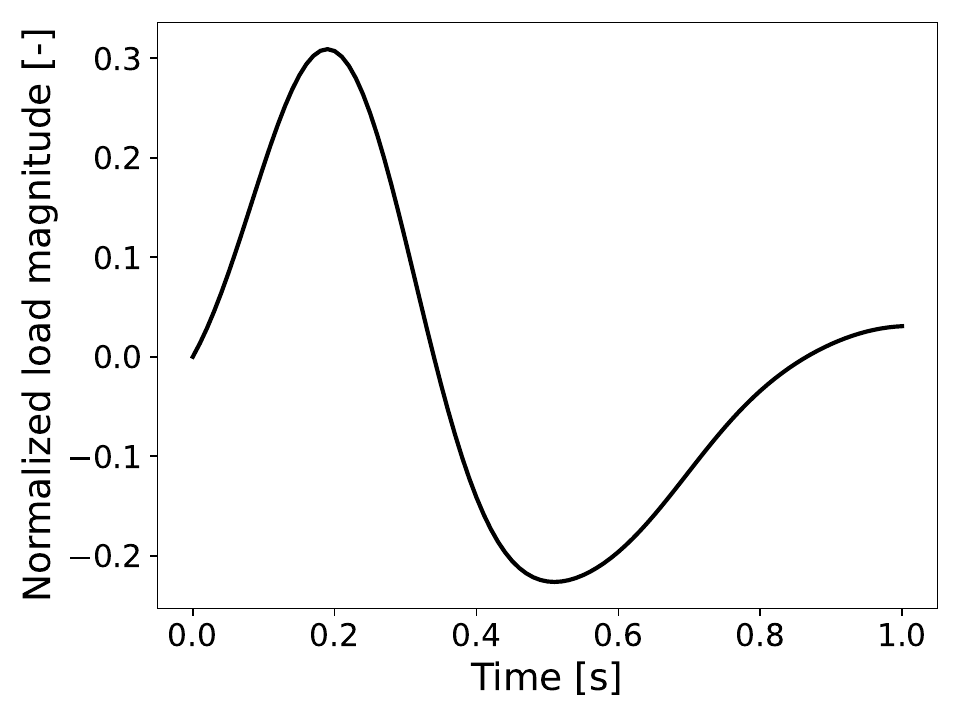}
        \label{lstm_mag}}
    \end{minipage} &
    \begin{minipage}[c]{.5\textwidth}
       \centering 
        \subfloat[(c) LSTM, worst case]{\includegraphics[trim={1cm 1.5cm 1cm 5.7cm},clip,width=\textwidth]{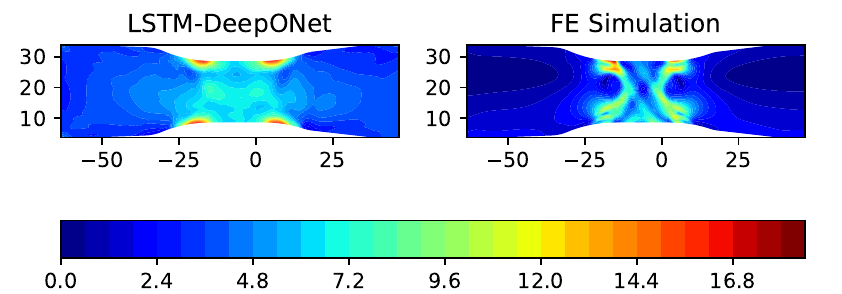}
        \label{lstm_worst}}
    \end{minipage} \\
    \end{tabular}
    \caption{Normalized load magnitudes and Mises stress contours for the worst-case prediction by the three DeepONet models. }
    \label{plastic_contour_results_worst}
\end{figure}

\begin{figure}[h!] 
    \centering
     \subfloat[FNN-DeepONet]{
         \includegraphics[trim={0cm 0cm 0cm 0cm},clip,width=0.3\textwidth]{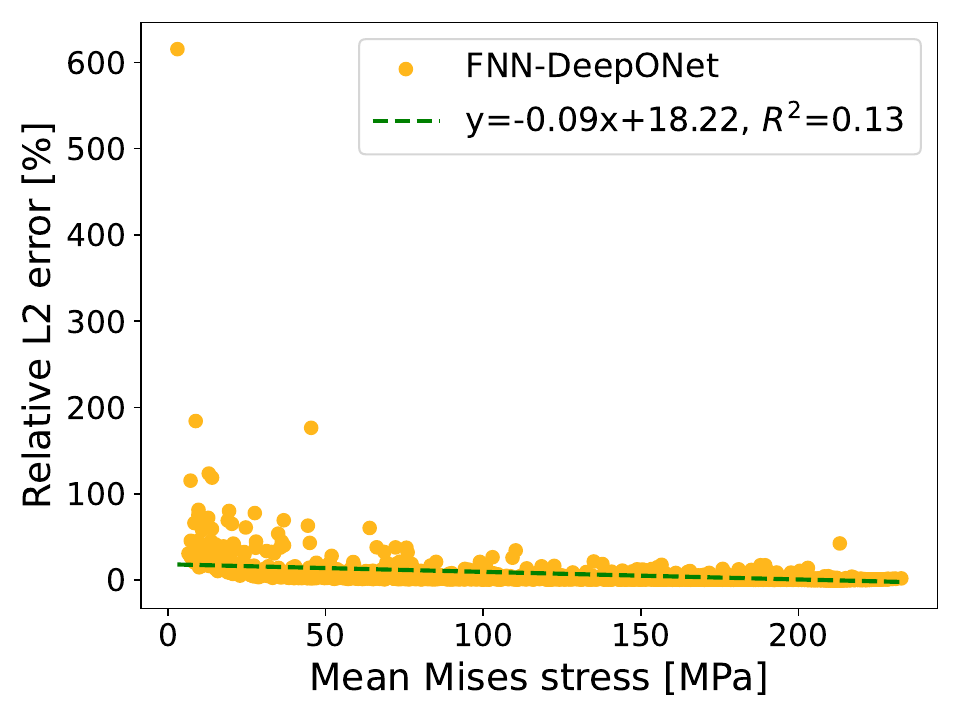}
         \label{e0}
     }
     \subfloat[GRU-DeepONet]{
         \includegraphics[trim={0cm 0cm 0cm 0cm},clip,width=0.3\textwidth]{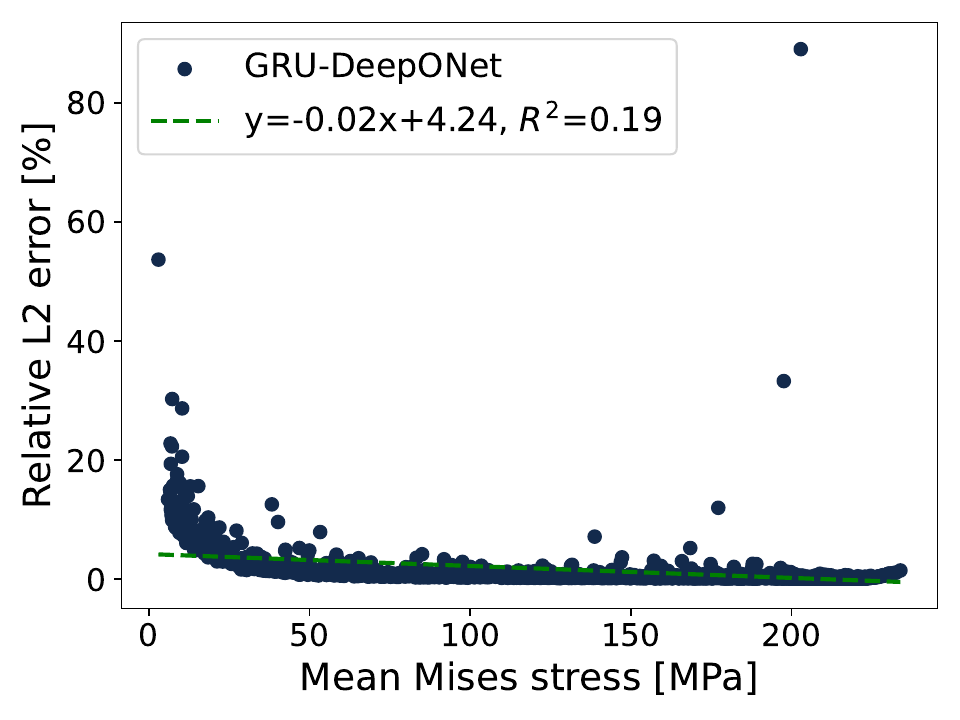}
         \label{e1}
     }
     \subfloat[LSTM-DeepONet]{
         \includegraphics[trim={0cm 0cm 0cm 0cm},clip,width=0.3\textwidth]{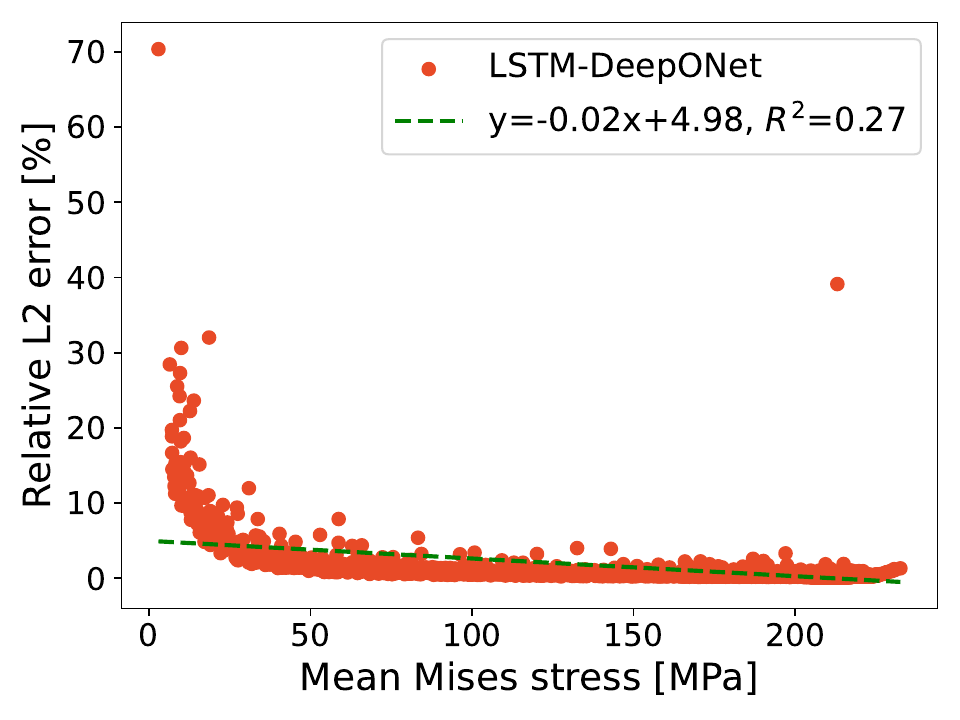}
         \label{e2}
     }
    \caption{Scatter plots of model prediction error versus the mean Mises stress over the domain as computed from the FE simulation. A linear curve fit was included in each subplot and the expression of each line is included in the legend.}
    \label{error_corr}
\end{figure}

For this more complex case, the worst-case predictions deserve special attention. To this end, the normalized load magnitudes and the von Mises stress contours for the worst predictions of each model are shown in \fref{plastic_contour_results_worst}. To further elucidate the relationship between prediction error and the mean stress magnitude, scatter plots are provided in \fref{error_corr}. 

With this more challenging problem, increasing the number of data points generally improves the performance of the models, with the best prediction accuracy achieved using 80\% of data in training for all three models. Again, the FNN-DeepONet shows the worst performance, with GRU-DeepONet providing the highest accuracy. Similar to the observations made in \sref{sec:ht_res}, the GRU- and LSTM-based models share comparable prediction accuracy, and the GRU-DeepONet demonstrated consistently accurate predictions in the 5-fold cross-validation, indicating that the results are repeatable. Both studies also show that the GRU- and LSTM-based models don't suffer from overfitting. The training time of the GRU model is 14.2\% faster than the LSTM model. Once trained, all three models can infer the full-field solution at a speed at least 200 times faster than FE simulations. \tref{plasticity_model_results} shows that the GRU- and LSTM-DeepONets are about 2.5 times more accurate than the FNN-DeepONet, at the expense of a 1.5 times longer training time. However, unlike the simple case in \sref{sec:ht_res}, we see that the model predictions for the plasticity case have significant outliers, with worst-case errors as high as 615\%. The presence of outliers is also evident in \fref{plastic_err_hist}, where there were at least 3.4\% of cases whose prediction error exceeds 5\%, despite having an overall $R^2$ value of over 0.995. 

We purposefully did not include the worst-case contours in \fref{plastic_contour_results} and defer that to \fref{plastic_contour_results_worst}. Excluding the outliers, we see from \fref{plastic_contour_results} that all three DeepONet models can accurately predict the stress contours and stress concentration points in the dog-bone specimen up to 90$^{th}$ percentile of the prediction error. As we get closer to the outliers (99$^{th}$ percentile), we see that the FNN-DeepONet predictions completely missed the two stress concentration points in the resulting contour, while the GRU- and LSTM-DeepONets were able to capture the location of the stress concentration points and predict the stress magnitude relatively accurately. These findings prove that the proposed DeepONet architectures with GRU and LSTM branches can effectively encode complex, time-dependent loading histories and perform better than the traditional FNN-based DeepONet. The spatial distribution of the predicted stress field closely resembles the FE ground truth, which indicates that the proposed DeepONets combine the temporal encoding capability of the GRU and LSTM structures with the spatial encoding capability of the DeepONet architecture to improve prediction accuracy.

From the worst-case prediction contours shown in \fref{plastic_contour_results_worst}, we see that in the worst-case, none of the three models were able to capture the location of the stress concentration points and the stress magnitude. With the exception of the worst case for the GRU model, the other two worst cases can be characterized by low stress magnitudes over the entire domain (as calculated from FE simulations). Additionally, from \fref{plastic_contour_results}, it is worth noting that the mean stress magnitudes for the higher error cases (99$^{th}$ percentile) seem to be lower than the better-performing cases as well. These observations prompted us to investigate the relation between DeepONet prediction errors and the ground truth mean stress magnitude, which is reported in \fref{error_corr}. From the results, we did notice a generally decreasing trend in the prediction error as the mean stress magnitude increases, with most of the poor predictions concentrated in cases with low ground-truth stress magnitudes. This observation is reasonable since the mean squared error is used as the loss function of training, therefore, regions/cases with small stress magnitudes are going to contribute less to the overall loss function, hence they do not get improved efficiently as the model trains. This could be mitigated by introducing a relative loss function definition for training and will be investigated in future works.

Similar to the transient heat transfer example in \sref{sec:ht_res}, the GRU- and LSTM models delivered similar prediction accuracy in this example. With the 14.2\% faster training time, the GRU model again emerges as a more computationally efficient choice in the two DeepONet architectures proposed in this work.

\section{Conclusions, limitations, and future work}
\label{sec:conc}
The classical data-driven DeepONet framework with the feed-forward neural network (FNN) in the branch and trunk is effective but ignores causal relations in the input data if the inputs to the branch network are time-dependent. Realizing this limitation, we introduced two sequential DeepONet architectures with advanced, recurrent neural networks of LSTM and GRU types in the branch, and this is the most significant and novel scientific contribution of this work. By introducing sequential learning models in the branch network of the DeepONet structure, we combined the powerful temporal encoding capability of the GRU and LSTM structure in the branch with the spatial encoding capability of the DeepONet architecture, which allows the accurate prediction of full-field solution contours given a time-dependent input load. We have then focused on learning full-field temperature and stress solutions in the two highly nonlinear practical applications of thermal and mechanical types with complex random loading histories. In both cases, GRU- and LSTM-DeepONets provided significantly more accurate predictions than classical DeepONet, proving that sequential learning methods in the branch of DeepONet can universally and effectively encode loading histories regardless of the underlying physics of the problem. DeepONets with a sequential branch network were able to half the average error among all testing samples compared to FNN-DeepONet. The difference was even more profound for the plasticity example, where GRU- and LSTM-DeepONets reduced the prediction errors by 2.5 times and accurately predict the Mises stress contour up to 90$^{th}$ percentile of prediction error. Through the two examples, we have shown that the proposed DeepONets can be adequately trained on the current high-end GPUs within a few hours. Moreover, once the DeepONets are properly trained, they can infer accurate full-field results at least two orders of magnitude faster than classical nonlinear numerical methods. Between the two proposed DeepONet architectures, the GRU-DeepONet has fewer trainable parameters compared to the LSTM-DeepONet, while it achieved similar prediction accuracy and trained faster during training. Therefore, it is recommended to use the GRU-DeepONet over classical FNN-DeepONet and LSTM-DeepONet for predicting the full-field solution given a time-dependent load. 

Although our current methodology yielded high accuracy, the proposed methods have certain limitations. Currently, only the end state of the time-dependent loading is predicted instead of the full evolution history. Secondly, many data in real-world applications are inherently unbalanced. For example, for real-world engineering structures, the majority of the load histories are centered around a certain design load, with only rare loading cases (such as sudden wind loads and earthquakes) that significantly exceed the design load level. This data imbalance was not accounted for in the data generation procedure. For both example problems, we uniformly sampled the inputs from for a wide range of load magnitudes. Further re-sampling treatments \cite{leng2022bi, leng2021loosely, leng2022cloud} should be employed to account for the imbalance effect of the training dataset.

With the improved accuracy afforded by the GRU and LSTM branch networks, the sequential DeepONet models can be used as an accurate and efficient surrogate model for FE simulations and can be used in high-fidelity multi-scale modeling, optimization, and design scenarios whenever a high number of forward evaluations with parametric histories are needed in many complex nonlinear and non-equilibrium applications and processes in engineering and science. The improved prediction accuracy compared to classical FNN-DeepONets also offered more confidence when deploying the model in practical engineering applications. Furthermore, the weights and biases of the trained sequential DeepONets can be transported to laptops and even edge computing devices and used for inference without the use of GPUs or even modeling tools for instant predictions in many online control scenarios. As the capabilities and memory of GPU hardware rapidly increase, in our future work, we will modify the current DeepONet architectures for learning full solution fields from three-dimensional modeling domains as well as predicting the complete time history of the solution contours.

\section*{Replication of results}
The data and source code that support the findings of this study can be found at \url{https://github.com/Jasiuk-Research-Group}. \textcolor{red}{Note to editor and reviewers: the link above will be made public upon the publication of this manuscript. During the review period, the data and source code can be made available upon request to the corresponding author.}

\section*{Conflict of interest}
The authors declare that they have no conflict of interest.

\section*{Acknowledgements}
The authors would like to thank the National Center for Supercomputing Applications (NCSA) at the University of Illinois, and particularly its Research Consulting Directorate, the Industry Program, and the Center for Artificial Intelligence Innovation (CAII) for their support and hardware resources. This research is a part of the Delta research computing project, which is supported by the National Science Foundation (award OCI 2005572) and the State of Illinois, as well as the Illinois Computes program supported by the University of Illinois Urbana-Champaign and the University of Illinois System. Finally, the authors would like to thank Professors George Karniadakis, Lu Lu, and the Crunch team at Brown, whose original work with DeepONets inspired this research.

\section*{CRediT author contributions}
\textbf{Junyan He}: Methodology, Formal analysis, Investigation, Writing - Original Draft.
\textbf{Shashank Kushwaha}: Methodology, Formal analysis, Investigation, Writing - Original Draft.
\textbf{Jaewan Park}: Methodology, Investigation, Writing - Original Draft.
\textbf{Seid Koric}: Conceptualization, Methodology, Supervision, Resources, Writing - Original Draft, Funding Acquisition.
\textbf{Diab Abueidda}: Supervision, Writing - Review \& Editing.
 \textbf{Iwona Jasiuk}: Supervision, Writing - Review \& Editing.

\bibliographystyle{unsrtnat}
\setlength{\bibsep}{0.0pt}
{\scriptsize \bibliography{References.bib} }
\end{document}